\documentclass[showpacs,amssymb,floatfix]{revtex4}
\usepackage{graphicx}
\usepackage{dcolumn}
\usepackage{bm}
\begin{document}
\title{Gate controlled resonant widths in double-bend waveguides: Bound states in the continuum}
\author{Almas F. Sadreev$^1$, Dmitrii N. Maksimov$^1$, and Artem S. Pilipchuk$^{1,2}$,}
\address{$^1$ Kirensky Institute of Physics Siberian Branch of
Russian Academy of Sciences, 660036, Krasnoyarsk, Russia}
\address{$^2$ Siberian Federal University, 660080 Krasnoyarsk,
Russia}
\date{\today}
\begin{abstract}
We consider quantum transmission through double-bend $\Pi$- and $Z$-shaped
waveguides controlled by the finger gate potential. Using the effective non-Hermitian
Hamiltonian approach we explain the resonances in transmission.
We show a difference in transmission in the short waveguides that is the result of
different chirality in $Z$ and $\Pi$ waveguides.
We show that the potential selectively affects the resonant widths
resulting in the occurrence of bound states in the continuum.
\end{abstract}
\maketitle

\section{introduction}

The effects of  bend discontinuities on  the
transmission   in double-bend quantum waveguides have been in the focus
of researches for long time \cite{Weisshaar,Wu,Vacek,Wang,Mekis,Carini,Shi,Ming}.
A Fano resonance was shown owing to  the  presence  of  a single
bend \cite{Sols,Shi,Ming}. The resonant picture  complicates in the double-bend waveguides
with the resonance widths  and positions dependent  on  the distance between bends
\cite{Goodnick,Kawamura}.
We believe that these resonances still are not properly understood.
One of the aims
of the present paper is to give a comprehensive description of resonant effects
in double-bend
quantum waveguides using the effective Hamiltonian approach \cite{Sokolov,Ingrid,Stockmann,Alhasid,Pichugin,SR}.
The central goal is however
to show that the finger gate potential selectively affects resonant widths
resulting in the occurrence of zero width resonances. That gives rise to
trapping of an electron between the bends or bound state in the continuum (BSC).

The phenomenon of localized state with discrete energy level embedded in the
continuum of extended states  originally considered in 1929 by von Neumann and Wigner
\cite{neumann} was long time  regarded as mathematical curiosity
because of certain spatially oscillating central symmetric
potentials. That situation cardinally changed when Ostrovsky {\it et al} \cite{ostrovsky}
and Friedrich and Wintgen \cite{friedrich} in framework of generic two-level Fano-Anderson model
formulated the
BSC  as a resonant state whose width tends to zero as, at least, one physical parameter varies
continuously. Localization of the resonant states of open system, i.e., the  BSC
is the result of full destructive interference of two resonance states which occurs
for crossing of eigenlevels of the closed system \cite{friedrich,volya,PRB}. That accompanied by avoiding crossing
of the resonant states one of which transforms into the trapped state with vanishing width
while the second resonant state acquires the maximal resonance width (superradiant state
\cite{ostrovsky,volya}). Recently the BSCs were considered in photonics
\cite{Shipman,Shabanov,photonic} that stimulated intensive experimental studies in electromagnetic
systems \cite{Lepetit,Segev,Longhi,Kivshar,Wei2,Lepetit1}. We address the
reader to Refs. \cite{Wei2,Wei}
to survey the current state of the art in the area of BSCs.

\begin{figure}
\includegraphics[height=6cm,width=6cm,clip=]{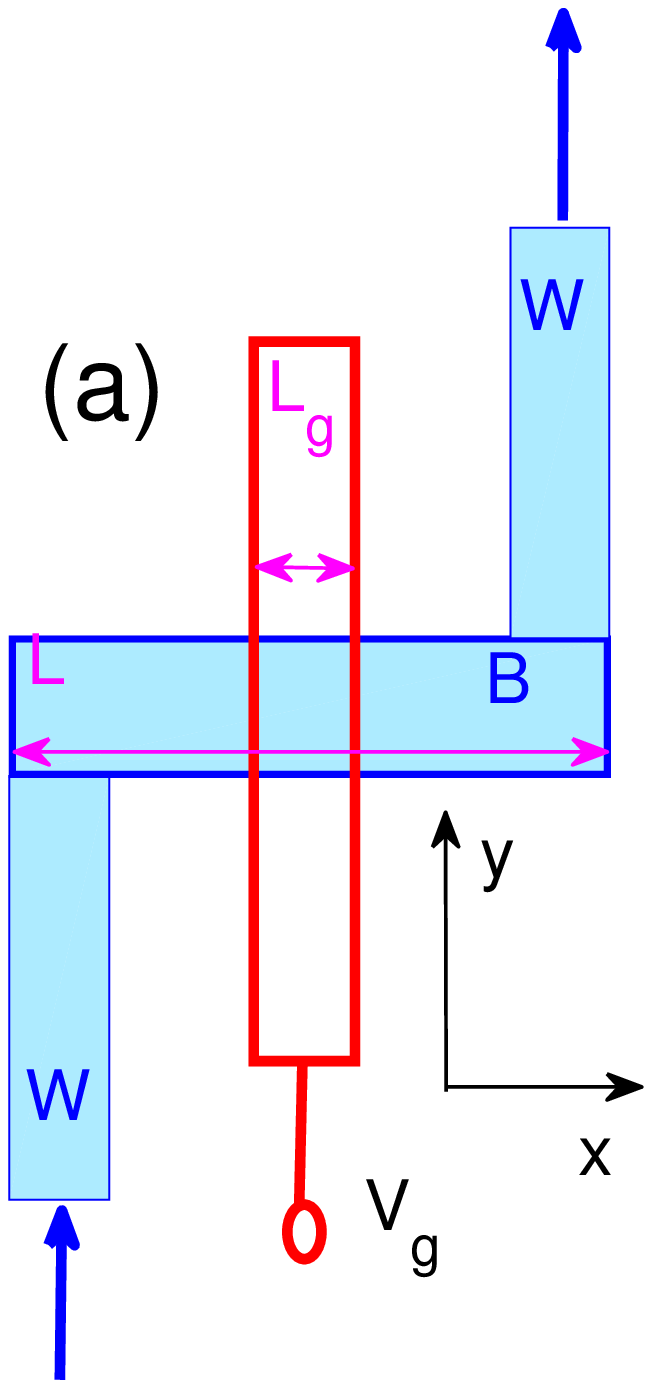}
\includegraphics[height=6.6cm,width=6cm,clip=]{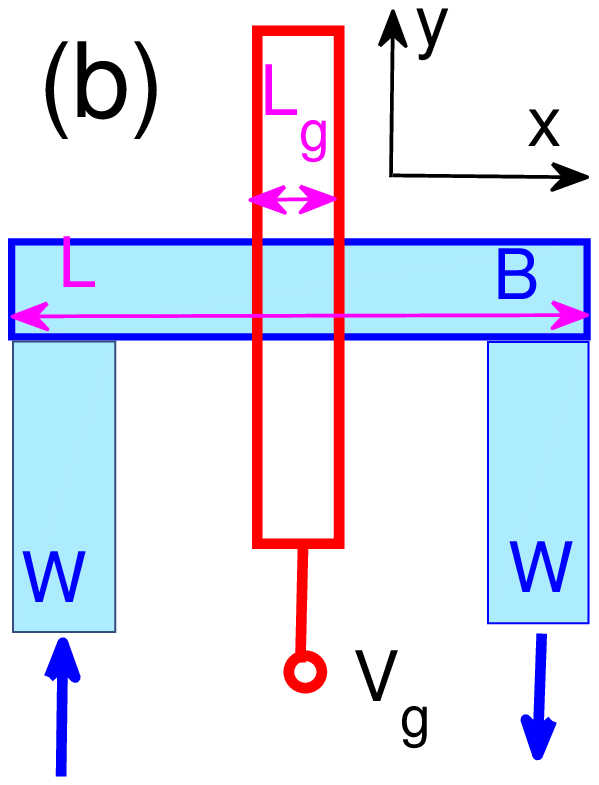}
\includegraphics[height=6cm,width=6cm,clip=]{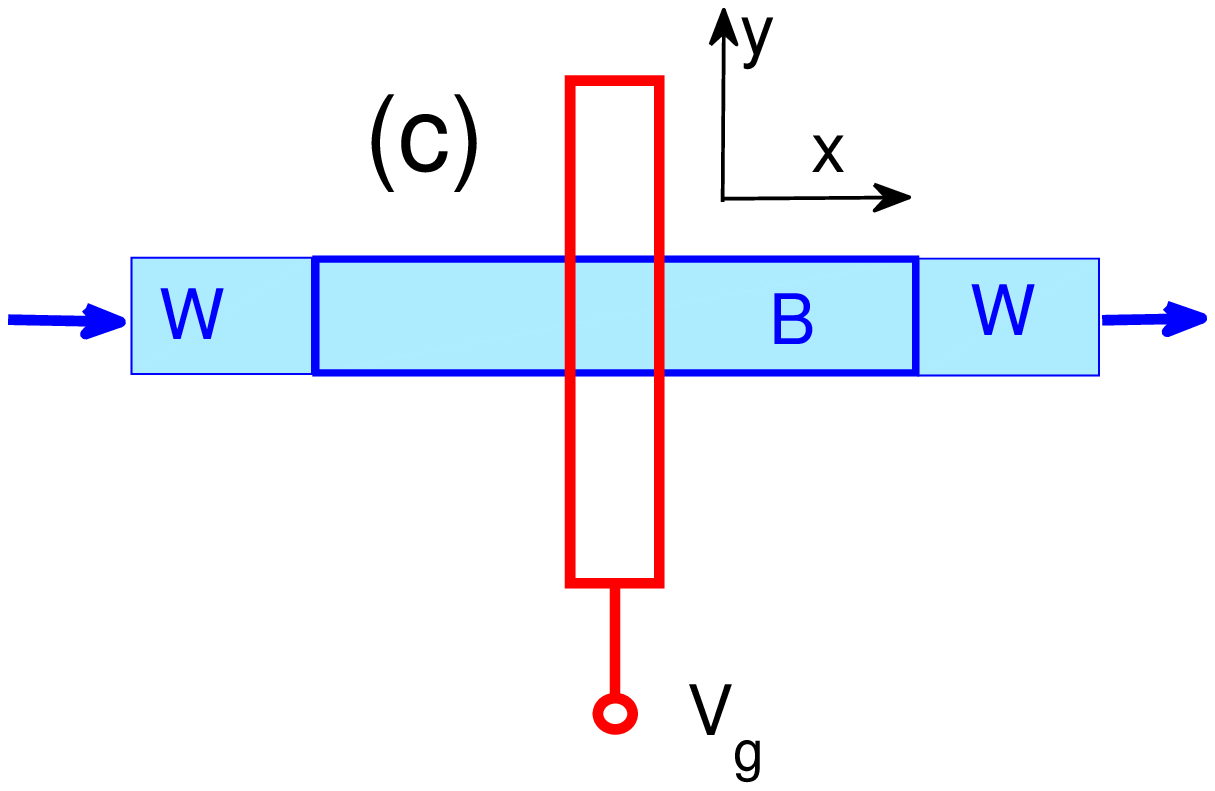}
\caption{(Color online) Double-bend waveguides, $Z$-shaped (a) and $\Pi$-shaped (b)
with  finger gate on the top. Arrows show electron flows.}
\label{fig1}
\end{figure}

In the present paper we consider wave transmission through two types
of the double-bend waveguides, namely $Z$- and $\Pi$-shaped quantum waveguides with a finger gate
positioned across to the wire as shown in Fig. \ref{fig1}. Each bend of the waveguides
has one transmission zero \cite{Weisshaar,Shi,Ming} at some energy $E_c$
for the first channel transmission. Therefore
the double-bend wire can be considered as a Fabry-Perot resonator (FPR) at the energy $E_c$. The FPR can trap an electron if the distance between the bends
is tuned to the integer number of half-wave lengths.
Such a tuning is however hardly experimentally plausible.
We propose to employ the finger gate potential
in order to effectively tune the distance between the bends.
We will show that multiple BSCs
related to integer numbers of half wavelengths between the bends occur for
positive potential. For a negative potential we will show the BSCs
which are close to those considered by Robnik \cite{Robnik}. Although in what follows we consider
the electron transmission through the quantum double-bend waveguides the consideration is
applicable to microwave transmission of TM waves owing to the equivalence between the Schr\"odinger
equation and the Maxwell equations for  planar electromagnetic waves. The dielectric slab inserted between
the bents of a waveguide could play the role of the finger gate potential.
\section{Transmission through double-bend waveguides}
The standard technique to reduce the resonant widths in electron transmission through a quantum dot
is to implement quantum point contacts.
In the present paper we consider a different approach to selectively control the coupling between
the inner states and the propagating states of the waveguides in the layouts shown in Fig. \ref{fig1} (a) and (b).
As shown in Fig. \ref{fig1} we split each double-bend waveguide into three parts. The inner part
outlined by bold blue line in Fig. \ref{fig1} and
denoted by "B" plays the role of a bridge between two semi-infinite directional waveguides denoted by "W".
The propagating states with the Fermi energy
\begin{equation}\label{Ep}
    E_F=\pi^2p^2+k_p^2, p=1,2,3,\ldots
    \end{equation}
in the waveguides are given by
\begin{equation}\label{propag}
 \psi_{\pm}(x,y)=\sqrt{\frac{1}{2\pi k_p}}\exp(\pm ik_px)\phi_p(y),
\end{equation}
where
\begin{equation}\label{phip}
    \psi_p(y)=\sqrt{2}\sin(\pi py).
\end{equation}
Here the Fermi energy is measured in terms of  $E_0=\frac{\hbar^2}{2m^{*}d^2}$, the coordinates are measured
in terms of the width of waveguides $d$  and the wave vector
$k_p$ is measured in terms of the inverse width of the waveguides.
The eigenfunctions of the inner part "B" are given by the following Schr\"odinger equation
\begin{equation}\label{HB}
\widehat{H}_B\psi_{mn}(x,y)=E_{mn}\psi_{mn}(x,y), ~~\widehat{H}_B=-\nabla^2+V_g(x),
\end{equation}
\begin{equation}\label{psimn}
\psi_{mn}(x,y)=\phi_m(x)\psi_n(y), ~~\psi_n(y)=\sqrt{2}\sin(\pi ny),
\end{equation}
\begin{equation}\label{Emn}
    E_{mn}=\epsilon_m(V_g)+\pi^2n^2,
\end{equation}
$\phi_m(x)$ and $\epsilon_m$ are the eigenfunctions and eigenenergies of a quantum particle in an
infinitely deep box of width $L$ with the finger gate potential $V_g(x)$
symmetrical in the $x$-axis. The exact analytical expression for the potential profile $V_g(x)$ was derived
in Ref. \cite{Davies}. If however
the finger gate is close enough to the 2DEG, the potential can be well approximated by a rectangular step-wise function
\cite{SadSher} with height $V_g$ and width $L_g$ equal to the width of the gate.
\begin{figure}
\includegraphics[height=5cm,width=6cm,clip=]{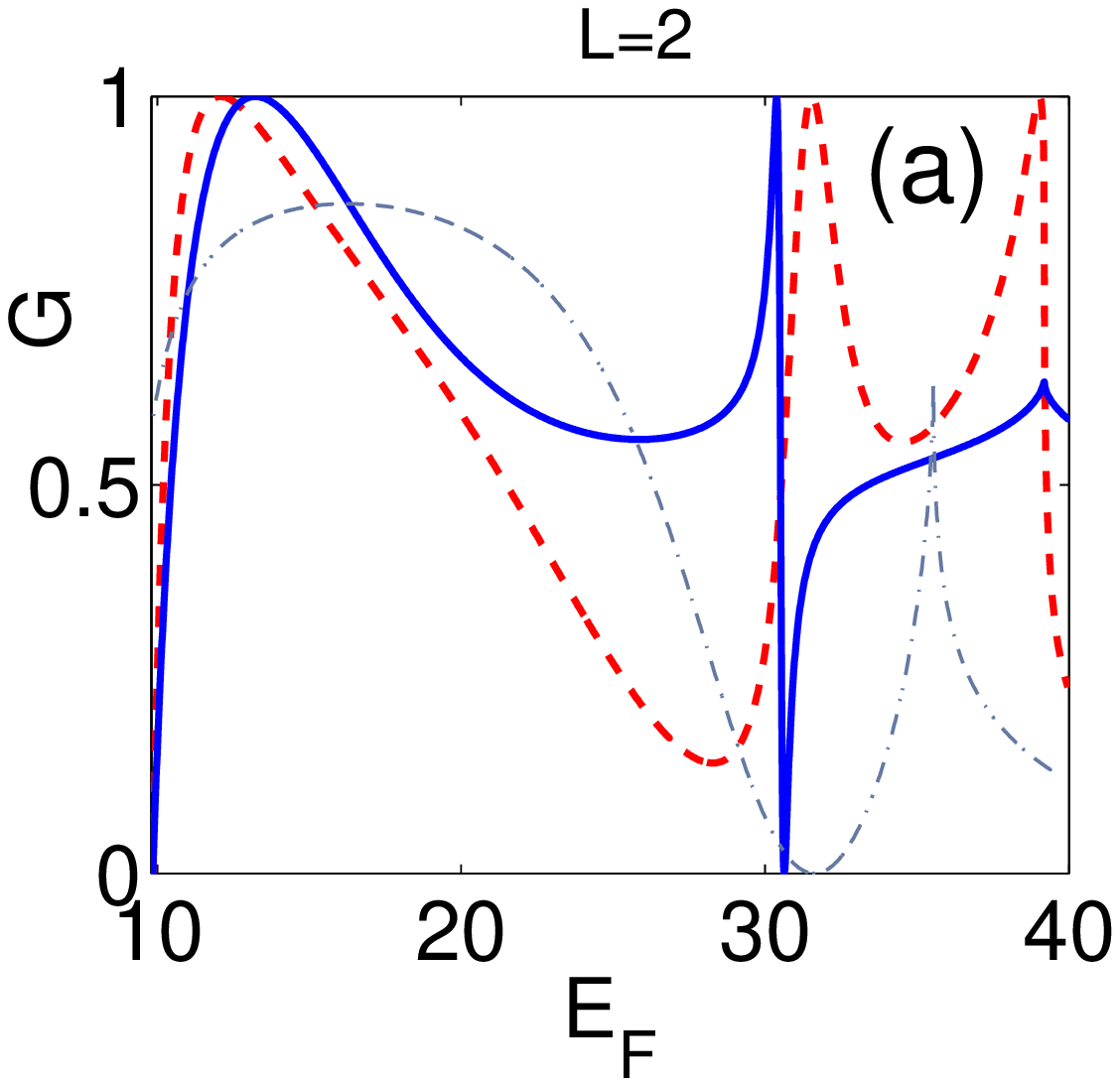}
\includegraphics[height=5cm,width=6cm,clip=]{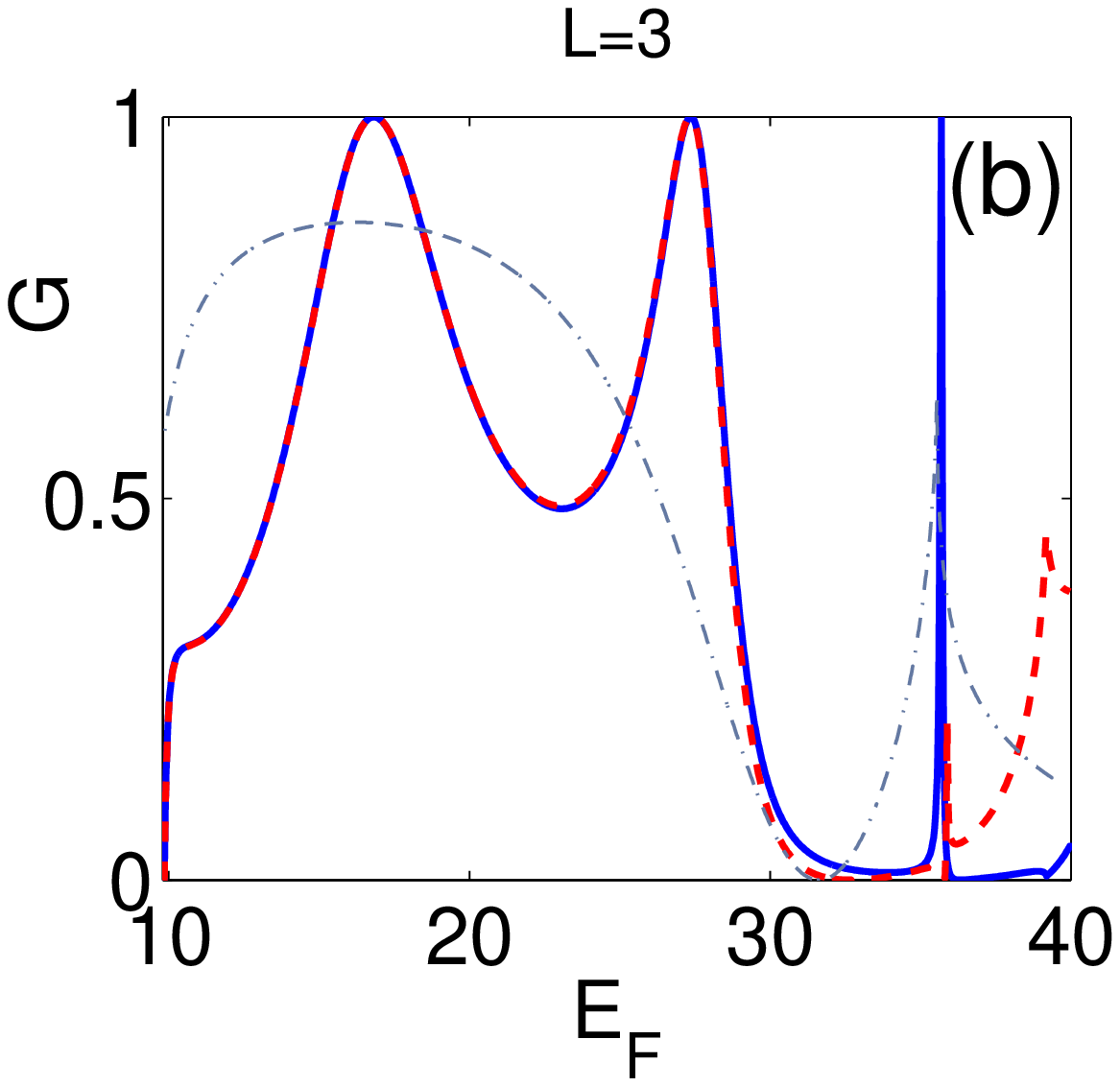}
\includegraphics[height=5cm,width=6cm,clip=]{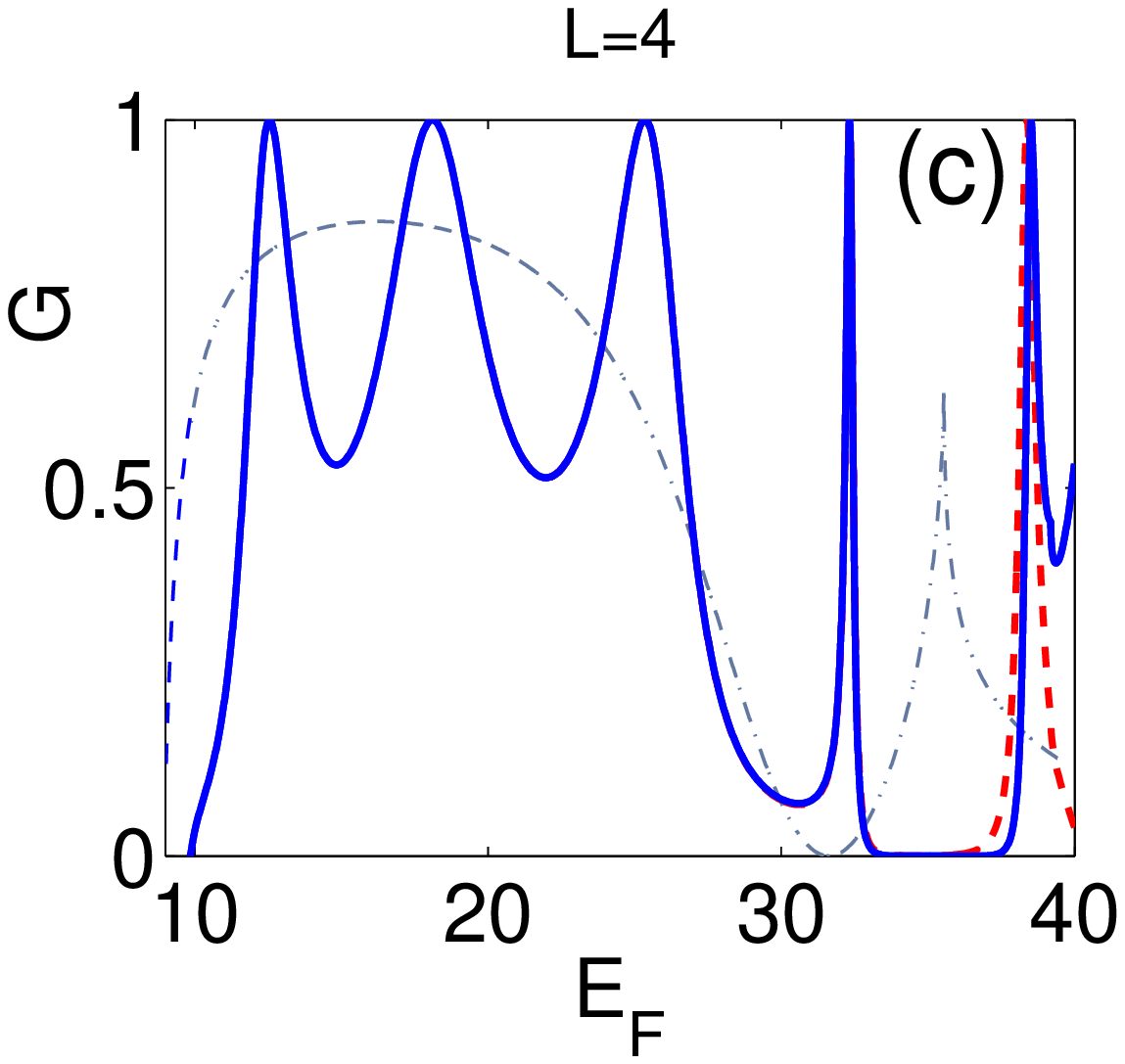}
\includegraphics[height=5cm,width=6cm,clip=]{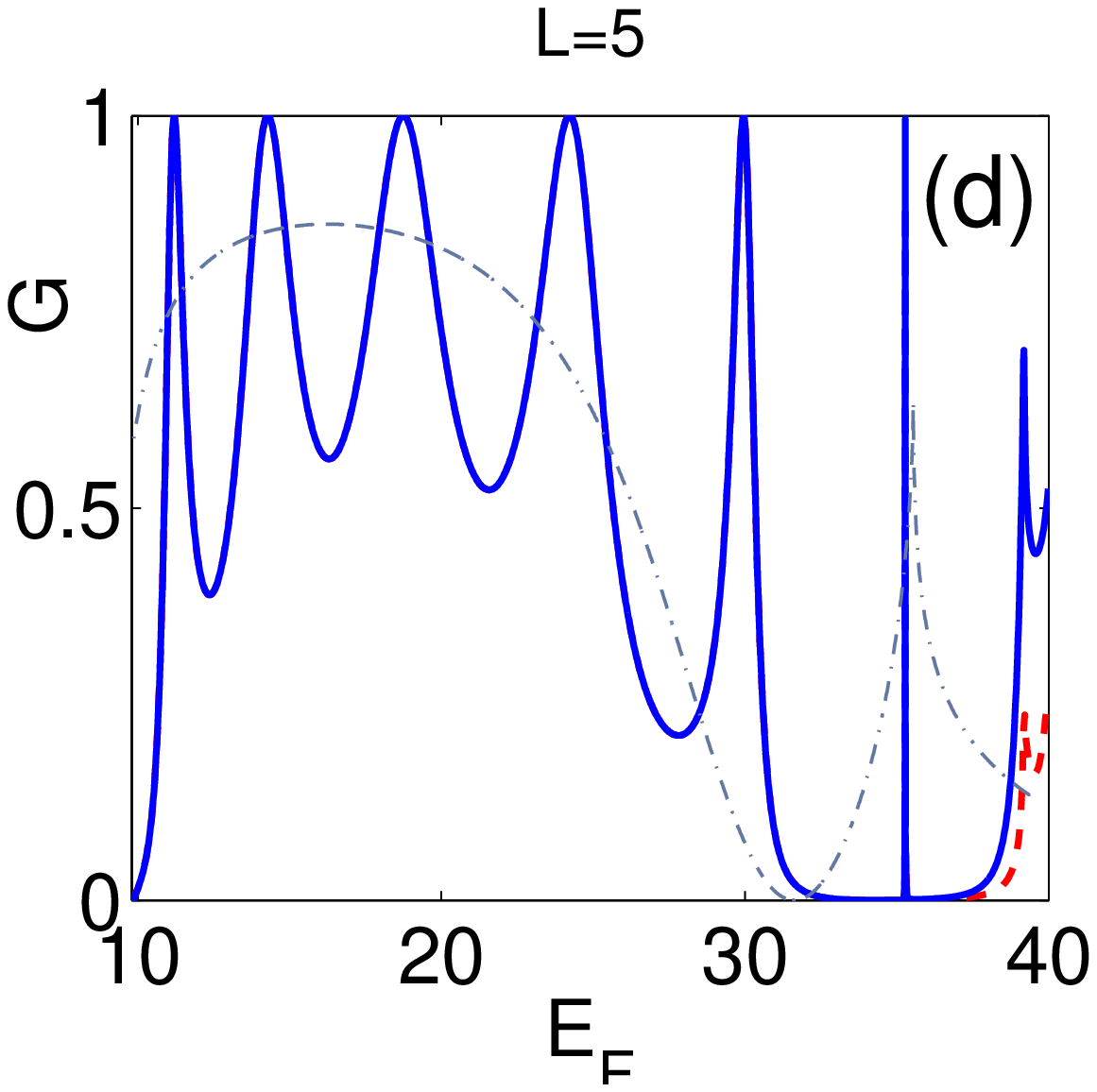}
\caption{(Color online) Conductance of double bent $Z$-shaped waveguide (red dash line) and
$\Pi$-shaped waveguide (blue solid line)
 for $V_g=0$ and for different lengths. The conductance of the single bent waveguide is
shown by green dot-dash line.}
\label{fig2}
\end{figure}

The conductance of double-bent waveguides could be calculated with the use
of the effective non-Hermitian Hamiltonian \cite{Sokolov,Ingrid,Stockmann,Alhasid,Pichugin,SR}
\begin{equation}\label{Heff}
    \widehat{H}_{eff}=\widehat{H}_B-i\pi\widehat{W}\widehat{W}^{\dagger}
\end{equation}
which is the result of projection of the total Hilbert space onto the inner space of the
bridge using the Feshbach technique \cite{feshbach,dittes}.
In this approach the waveguides are coupled to the bridge by
the matrix $\widehat{W}$ whose elements are the coupling constants of the
first channel propagating state (\ref{propag}) with the inner states (\ref{psimn})
calculated via overlapping integrals of the form \cite{Pichugin,SR,Savin}
\begin{equation}
\label{Wmn}
 W_{m,n}=\frac{1}{\sqrt{k_1}}\int_0^1 dx\psi_1(x)\frac{\partial\psi_{m,n}(x,y=0,1)}{\partial y}
 =\frac{2\pi n}{\sqrt{k_1}}\int_0^1 dx\psi_1(x)\phi_m(x).
\end{equation}
Then the transmission amplitude is given by \cite{Sokolov,Ingrid,Stockmann,Alhasid,SR}
\begin{equation}\label{T}
    t=-2\pi i\widehat{W}^{\dagger}\frac{1}{\widehat{H}_{eff}-E}\widehat{W}.
\end{equation}
\begin{figure}
\includegraphics[height=6.5cm,width=5.5cm,clip=]{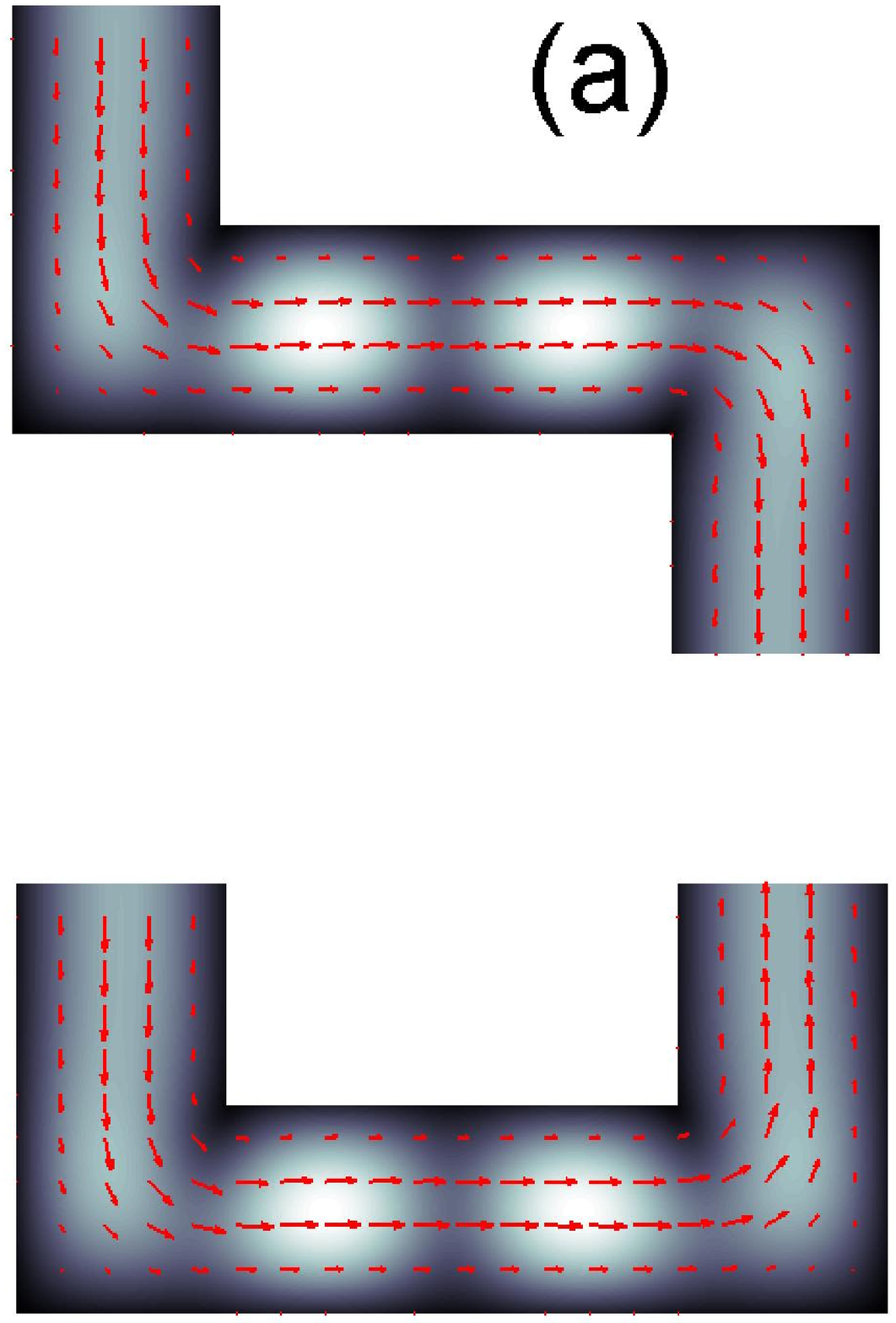}
\includegraphics[height=6.5cm,width=5.5cm,clip=]{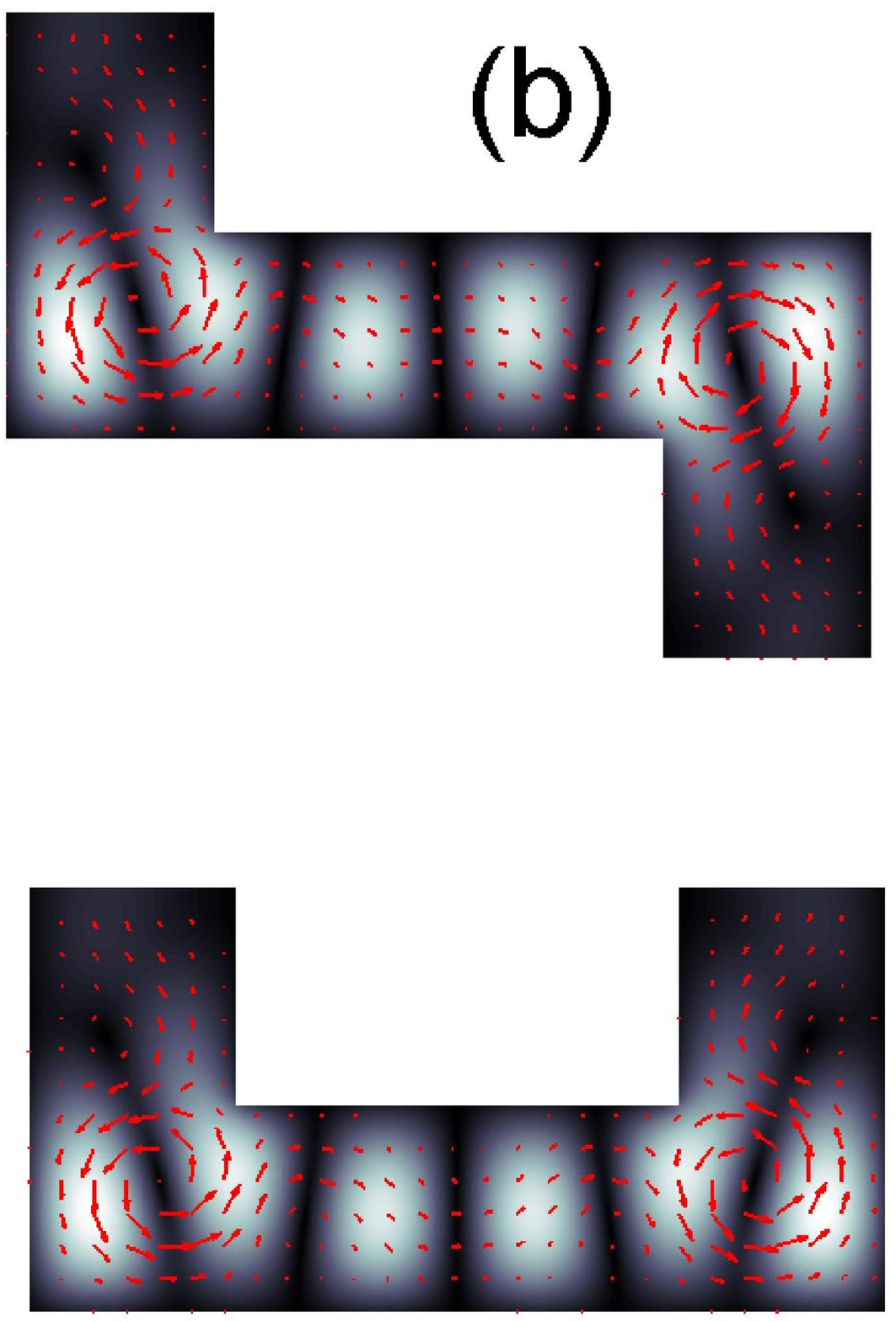}
\caption{(Color online) Scattering wave functions and current flows shown by red arrows
in double bent $Z$- and $\Pi$-shaped waveguides for energies $E_F=18.11$ (a) and $E_F=32.33$ (b) which correspond to the
second and fourth resonant peaks in Fig. \ref{fig2} (c) with $L=4$. }
\label{fig3}
\end{figure}
Computationally the method of the effective non-Hermitian Hamiltonian becomes efficient when
adapted to a discretized form \cite{Datta,SR} equivalent to the finite-difference approach
to the Schr\"odinger equation which is free from issues of poor convergence \cite{Pichugin}.
In Fig. \ref{fig2} we plot the conductance $G=|t|^2$ in the Z-shaped waveguide in comparison to the conductance
in the $\Pi$-shaped waveguide for $V_g=0$ for different lengths $L$.
First, one can observe resonant peaks which become narrower with the growth of $L$.
That observation was reported in Refs. \cite{Weisshaar,Wu,Wang,Mekis} however it did not
receive a clear explanation.
Second, below the threshold of the second channel $E_F=4\pi^2$ but $E_F>30$
we see sharp asymmetric
Fano resonances where the transmission can be either zero or unit depending on the type of the
waveguide.
For comparison we presented in Fig. \ref{fig2} the conductance of the single bend waveguide by dash-dot line which does
not exhibit resonances but has a transmission zero at energy $E_F\approx 31.62$.
Third, the conductance in the Z-shaped waveguide differs from the conductance in the $\Pi$-
shaped waveguide when the second channel of the bridge $B$ participates in the electron transmission.
The difference between waveguides tends to zero with the extension of the length  $L$ of the bridge.
The transmission resonances
for energies far below the second channel threshold $E<4\pi^2$ are typical for the transmission
through double barrier structure where peaks of the transmission correspond to standing waves between
the barriers. Fig. \ref{fig3} shows the scattering wave function
which indeed demonstrate standing waves at the transmission resonances.

\begin{figure}
\includegraphics[height=6cm,width=6cm,clip=]{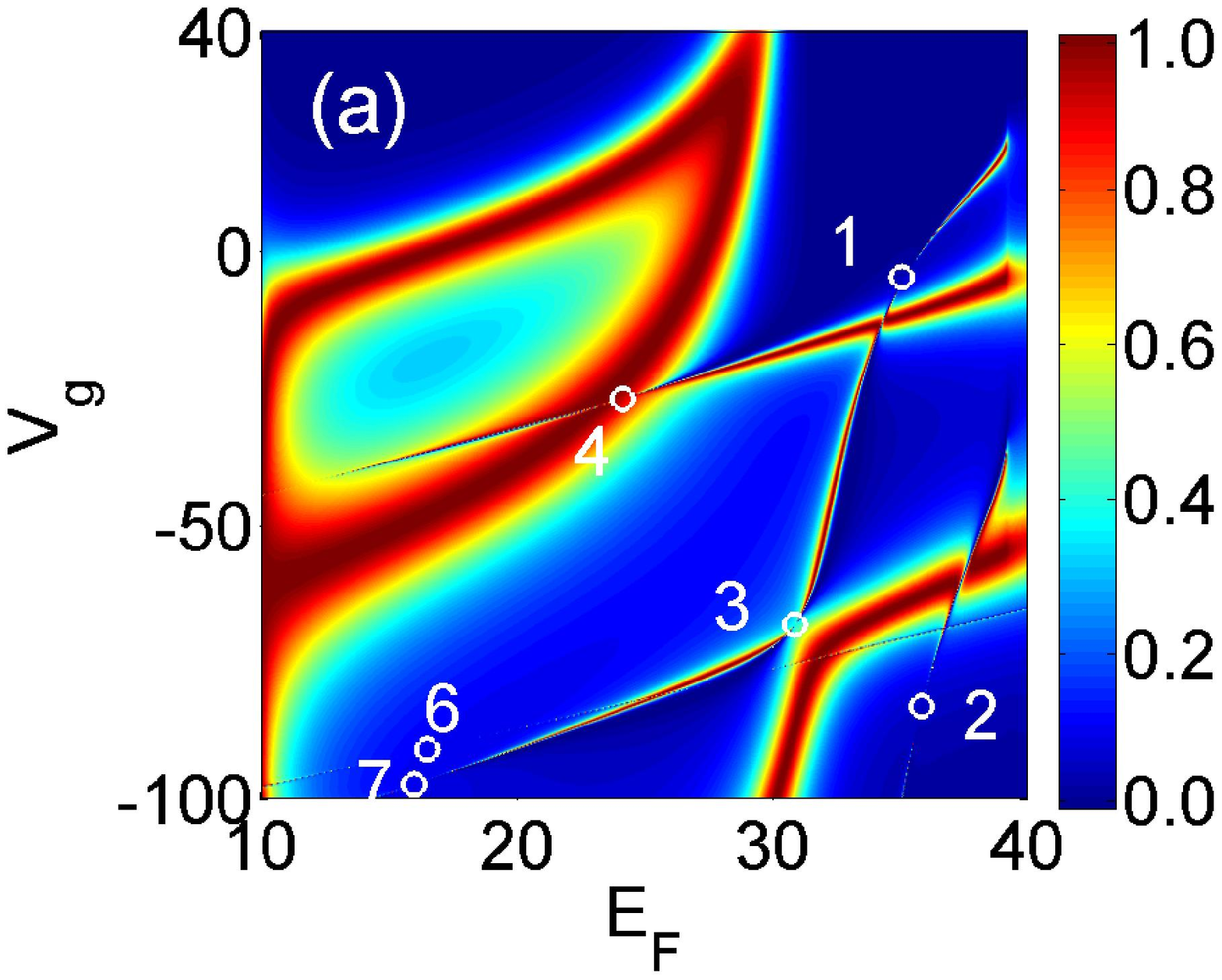}
\includegraphics[height=6cm,width=6cm,clip=]{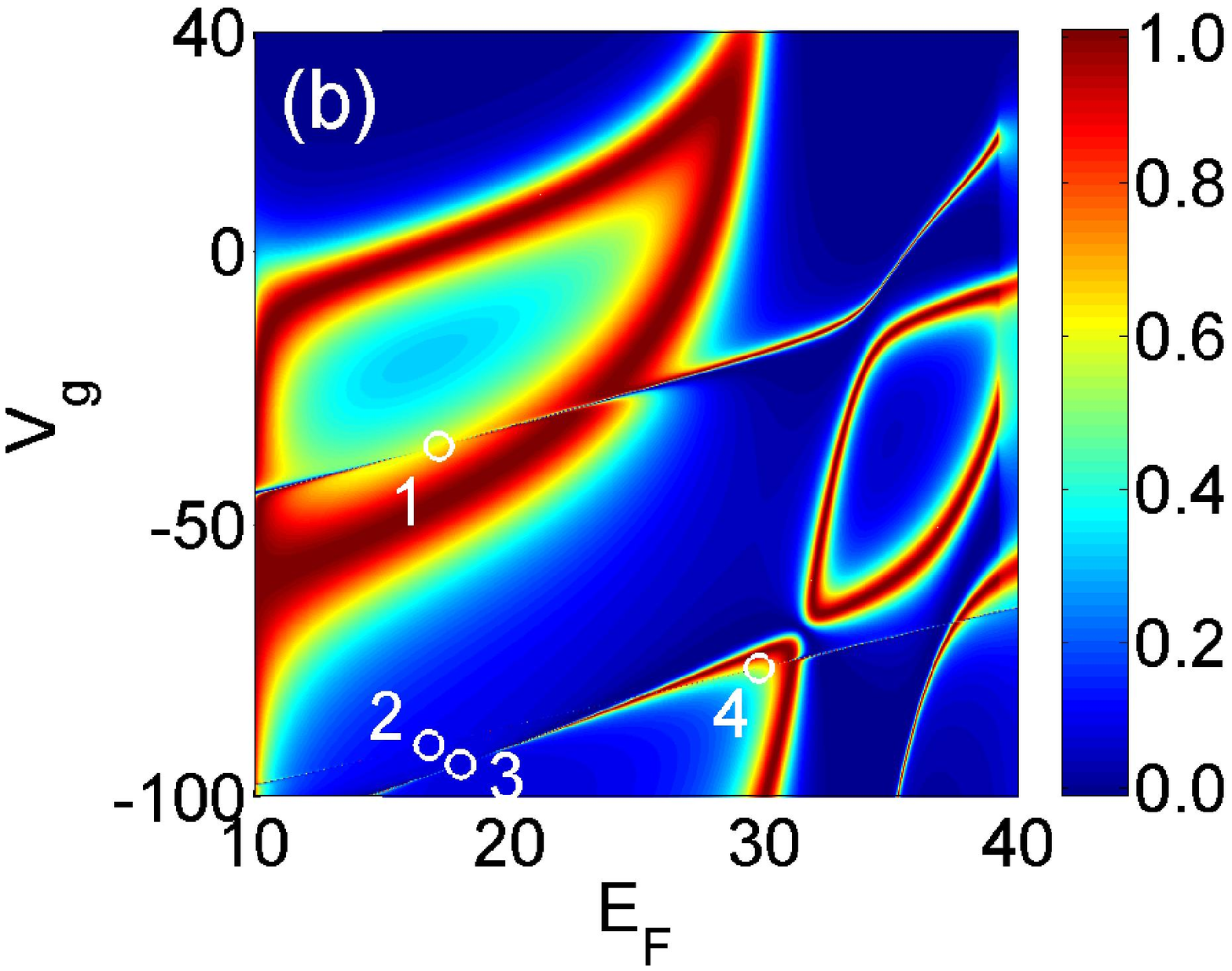}
\caption{(Color online) Conductance of double-bent waveguide vs Fermi energy and $V_g$ for $Z$- (a) and
and $\Pi$-shaped waveguides (b) for $L=3$. Open circles mark BSC. }\label{fig4}
\end{figure}

Figs. \ref{fig3} (a) and \ref{fig3} (b) show that
mainly the eigenfunctions $\psi_{m,1}(x,y)$ contribute to the scattering wave function
for long waveguides with $L=4$ with the Fermi energy
far from the second channel band edge. In the other words, the waveguides can be considered as
one-dimensional in which there is no difference between the chirality sequence of the bends.
The current flows demonstrate laminar regime respectively.
When the energy is approaching  the second channel edge $E=4\pi^2$ the contribution
of the second channel functions $\psi_{m,2}$ becomes relevant giving rise to vortical motion
as it was first observed by Berggren and Ji \cite{Karl&Ji} for electron transmission through a single bend
waveguide. The direction of current circulation at the vortices depends on the chirality of the bend.
Therefore the
$\Pi$-shaped waveguide with the bends of the same chirality have vortices with the same clockwise or counter-clockwise
current circulation while the $Z$-shaped waveguide with the bends of opposite chirality has vortices of opposite
current circulation as seen from Figs. \ref{fig3} (c) and \ref{fig3} (d).
As a result the current flows and respectively the transmission
depends on the type of the double-bent waveguide. Fig. \ref{fig4} shows that for the shortest bridge $L=2$
the current flow is vortical even for energies far from the second channel edge which gives rise
to a difference in conductance for the whole energy band as seen from Fig. \ref{fig2} (a).
In Fig. \ref{fig7} we show the first channel conductance
for both types of waveguides versus the Fermi energy and the finger gate potential for a long bridge $L=4$.
The waveguides show generally very similar conductance although with cardinally different pattern of the avoided
crossings. As will be shown below that is reflected in the number and types of the BSC which
denoted by open white circles in Fig. \ref{fig7}.
\begin{figure}
\includegraphics[height=7cm,width=5cm,clip=]{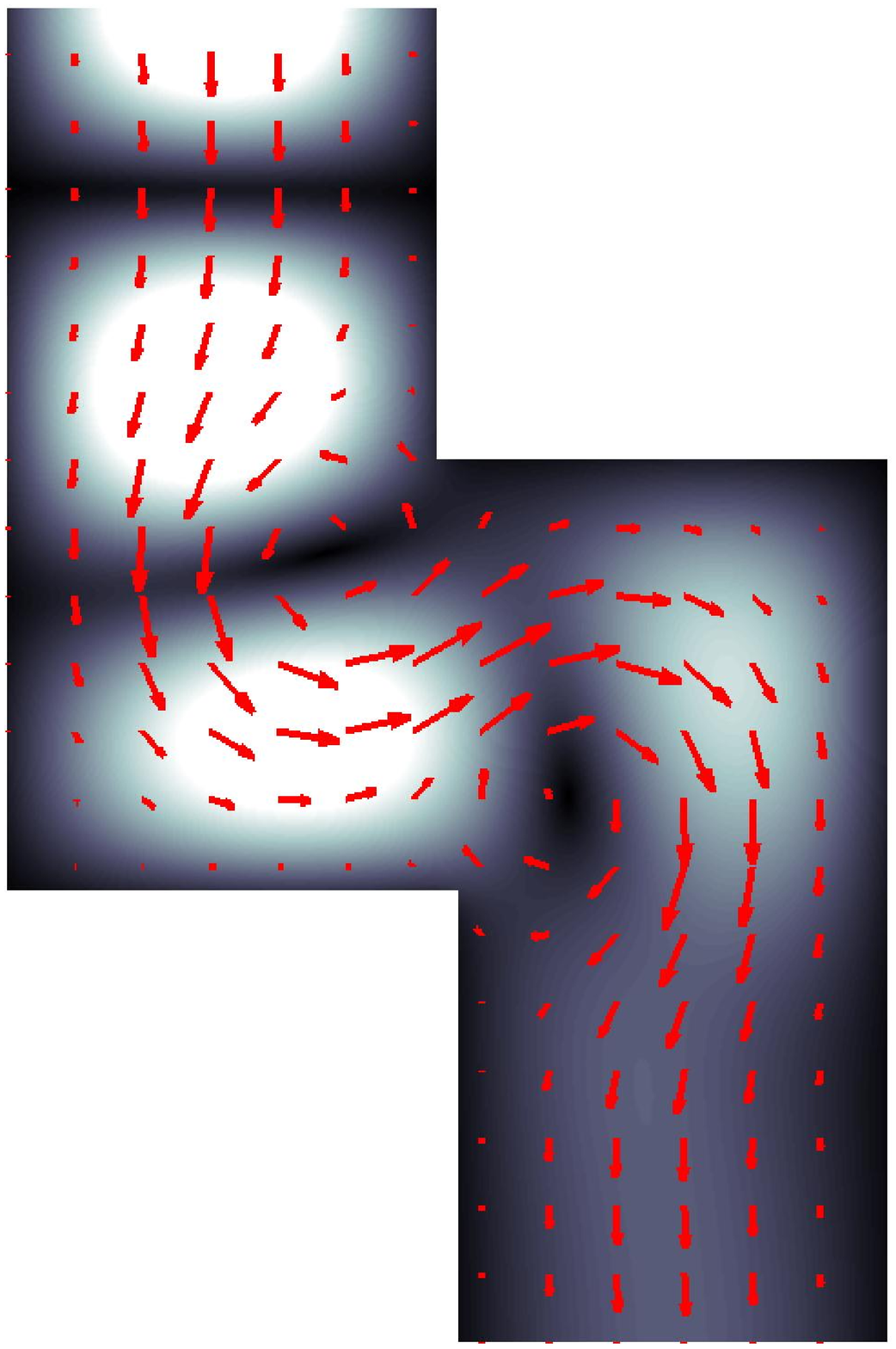}
\includegraphics[height=5cm,width=5cm,clip=]{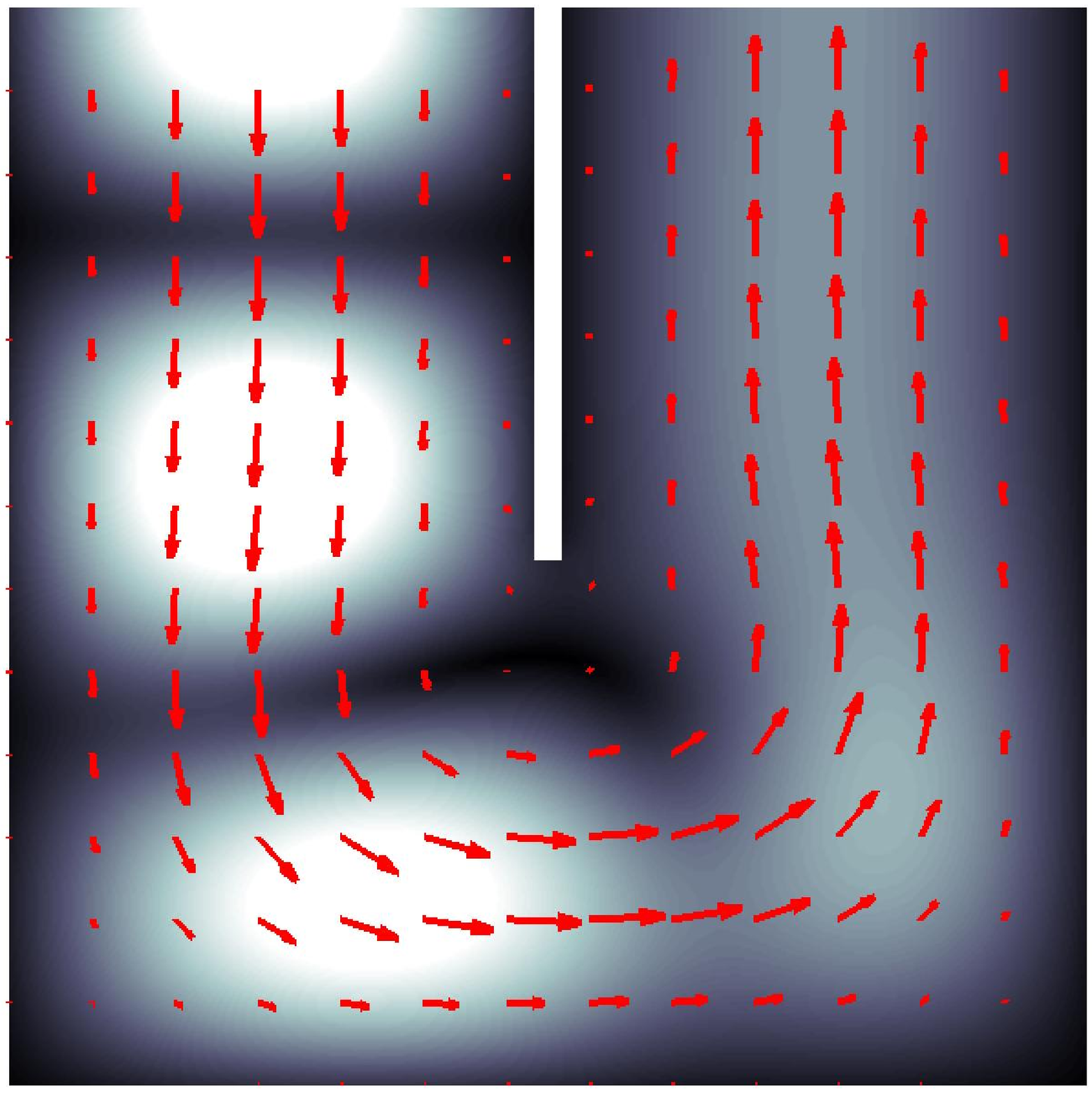}
\caption{(Color online) The same as in Fig. \ref{fig3} for $L=2$ and  $E=25$.}
\label{fig5}
\end{figure}

The resonant widths are controlled by the value of
the coupling matrix elements (\ref{Wmn}). Because the longitudinal  wave functions
$\phi_m(x)$ are normalized to have $L^{1/2}$ in the denominator we obtain that the resonant
widths are inversely proportional to $L$ according to Eq. (\ref{Heff}).
That observation explains the tendency of narrowing of resonant peaks
with the growth of the bridge length $L$ as shown in Fig. \ref{fig2}. Moreover, that
opens an opportunity to control the resonant widths by the potential of the finger gate.
This point has a key importance for the BSC and will
be considered in the next section.

\section{Bound states in the continuum}

\begin{figure}
\includegraphics[height=6cm,width=6cm,clip=]{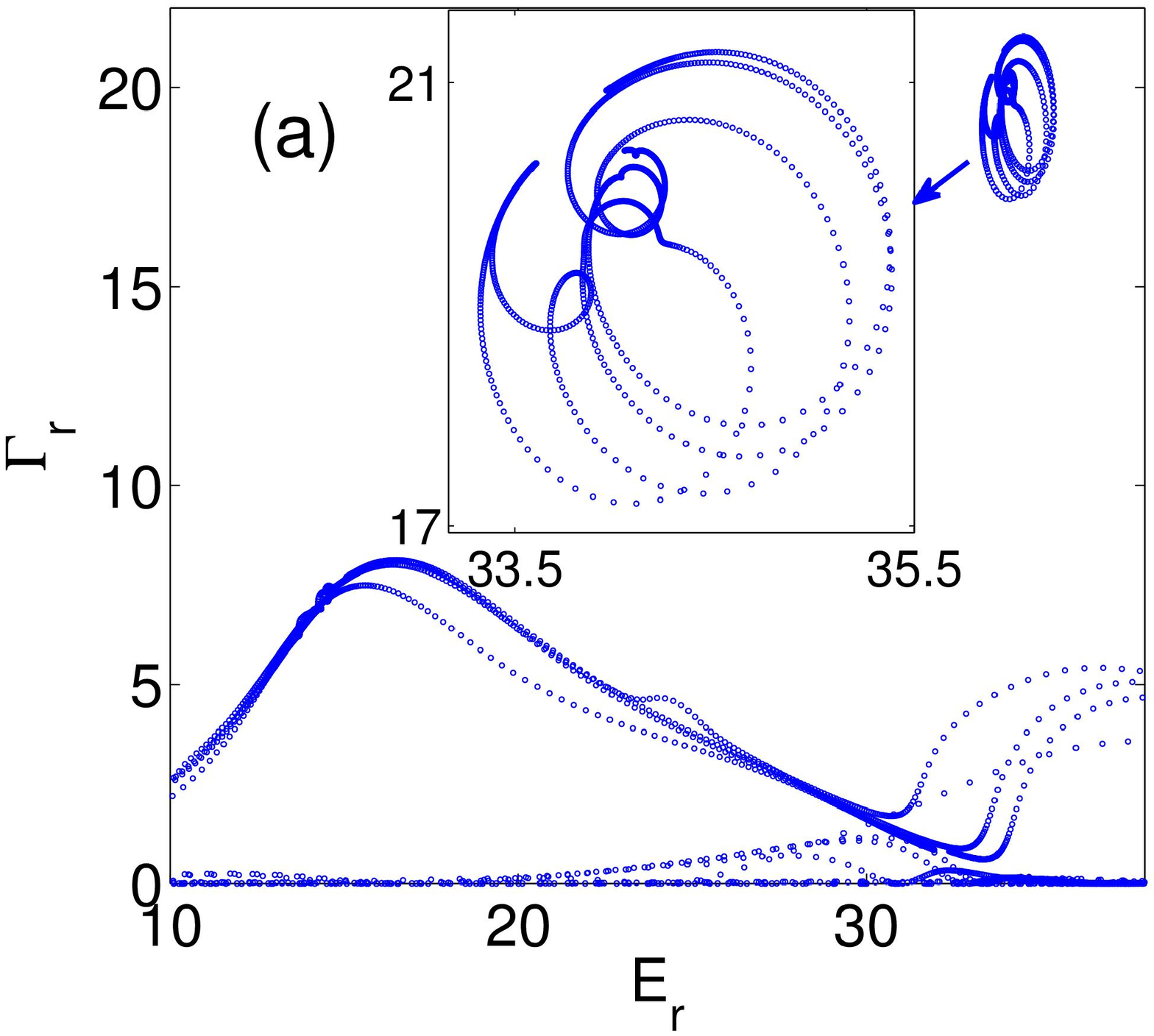}
\includegraphics[height=6cm,width=6cm,clip=]{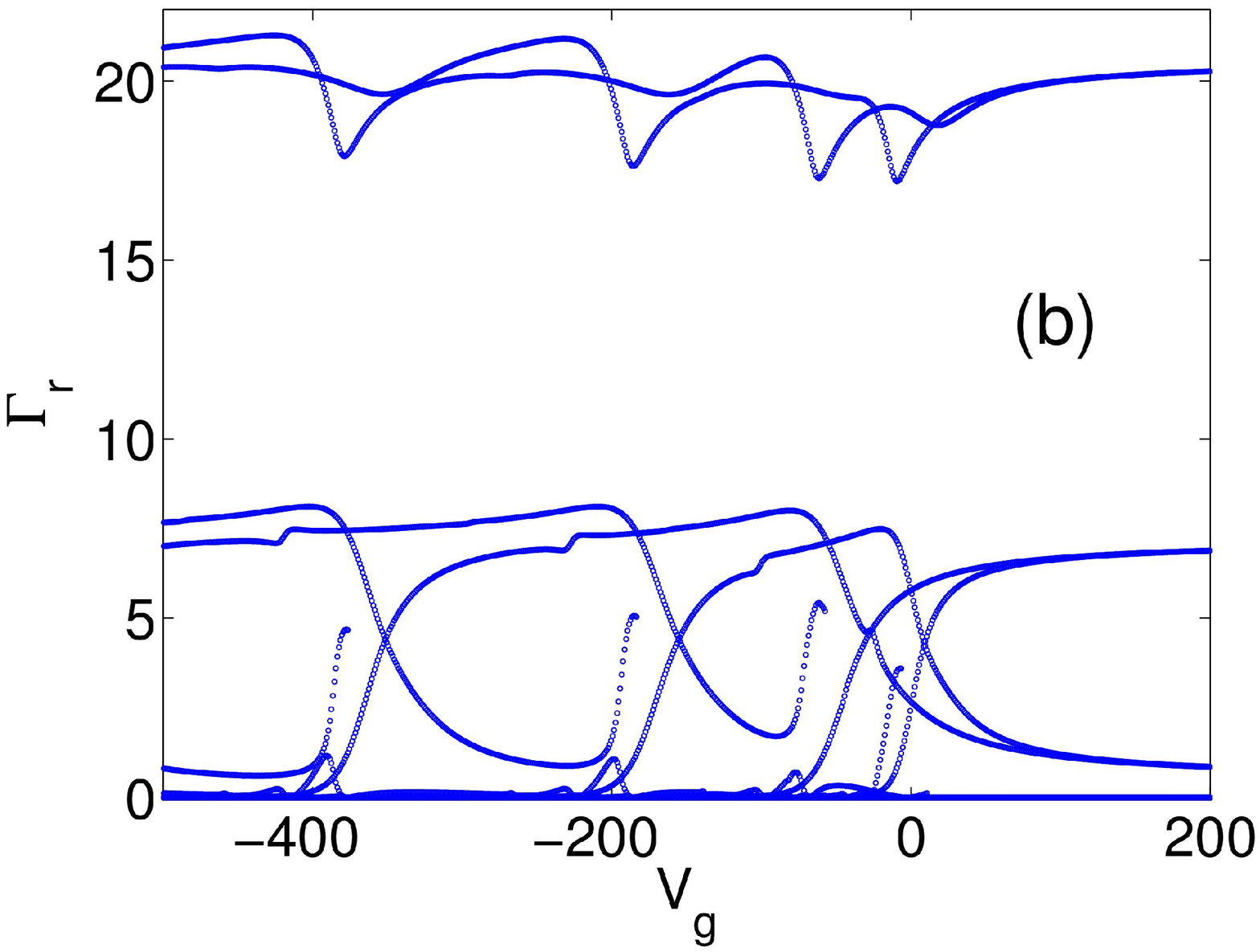}
\includegraphics[height=6cm,width=6cm,clip=]{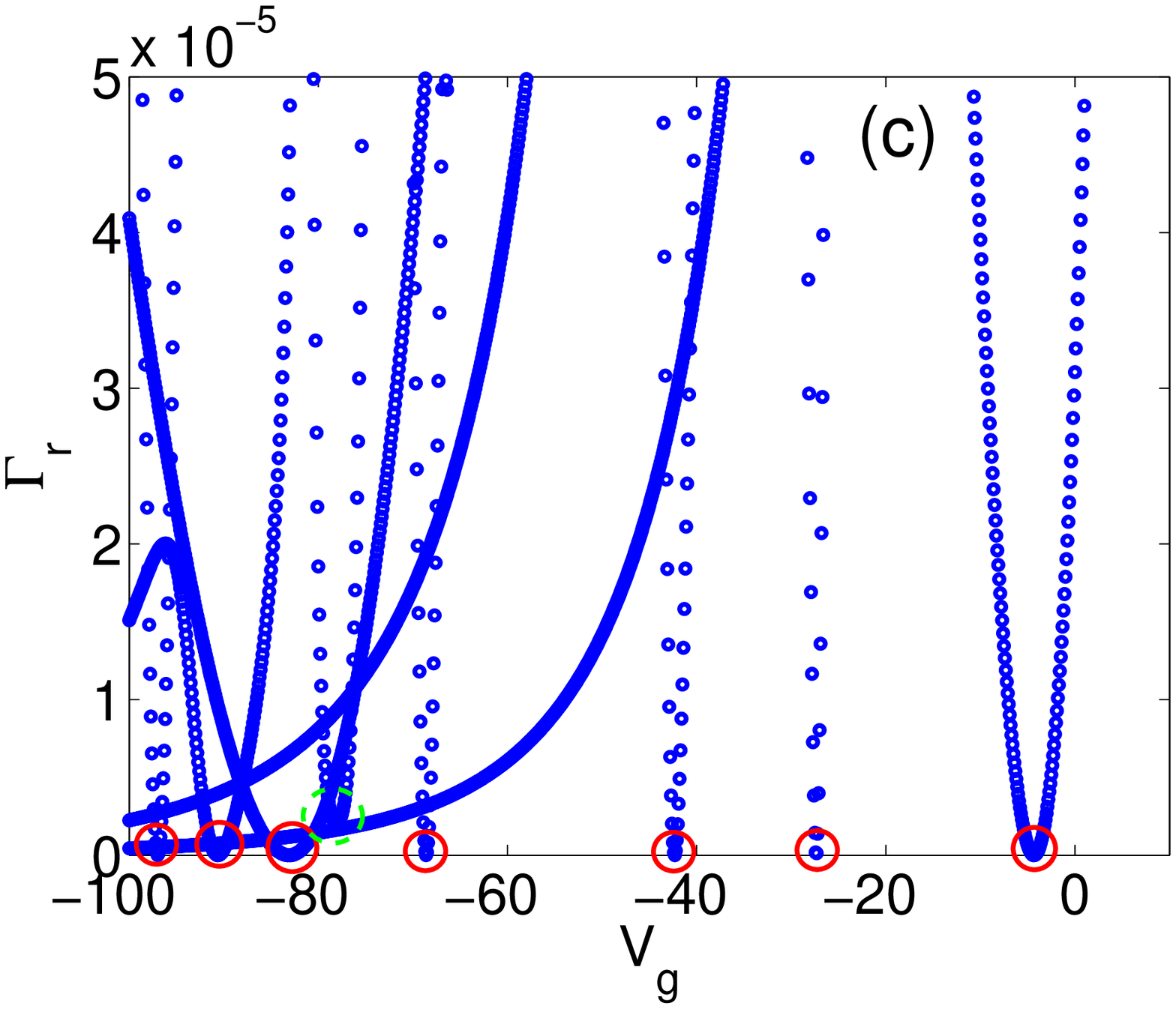}
\caption{(Color online) (a) Evolution of resonant positions and resonant widths with the finger gate potential
$V_g$ for the $Z$-shaped double bend waveguide for $E=25, L=3$. (b) The resonant widths vs $V_g$ zoomed in (c).}
\label{fig6}
\end{figure}

The complex eigenvalues $z$ of the effective Hamiltonian $\widehat{H}_{eff}$ have a simple
physical meaning \cite{Sokolov,Ingrid,Savin}. Their real parts $E_r=Re(z_)$ define the
positions of the resonances with
the resonance widths given by imaginary parts $\Gamma_r=-2Im(z)$.
In order to find the complex eigenvalues $z$ in case of noticeable radiation shifts
it is necessary to solve the fix point
equations for the resonance positions \cite{Ingrid} $E_r=Re[z(E_r)], \Gamma_r=Im[z(E_r)]$.
Resonant widths as functions of the finger gate potential $V_g$ show in
Fig. \ref{fig5} oscillating behavior because the longitudinal wave functions of the bridge $\phi_m(x)$  tend
to accumulate their nodal points on the waveguide-bridge interface
as the probability distribution is pushed from the central region of the bridge with  the growth of the finger gate potential
as shown in Fig. \ref{fig6} (a). As a result the overlapping integrals
\begin{figure}
\includegraphics[height=6cm,width=6cm,clip=]{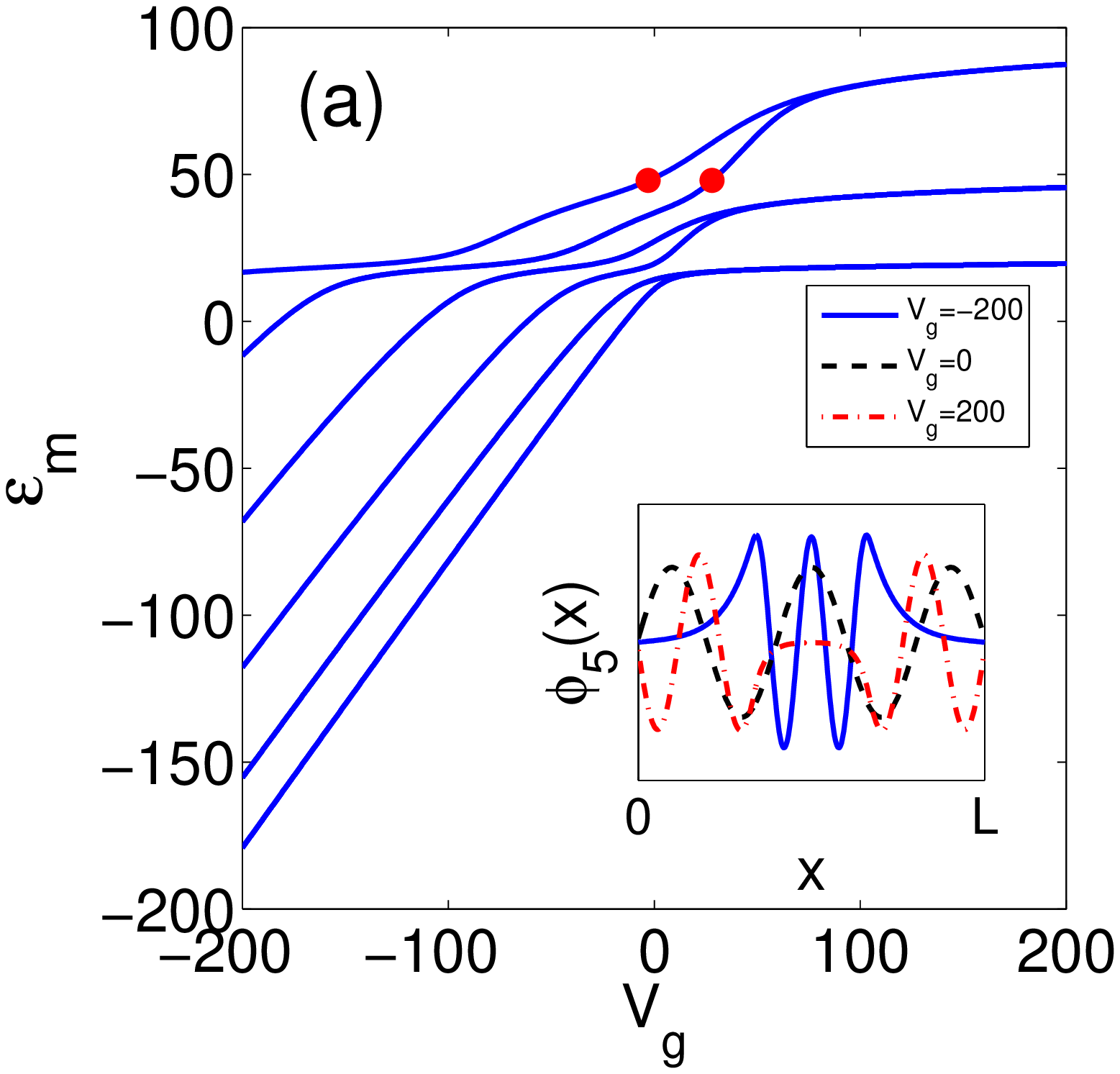}
\includegraphics[height=6cm,width=6cm,clip=]{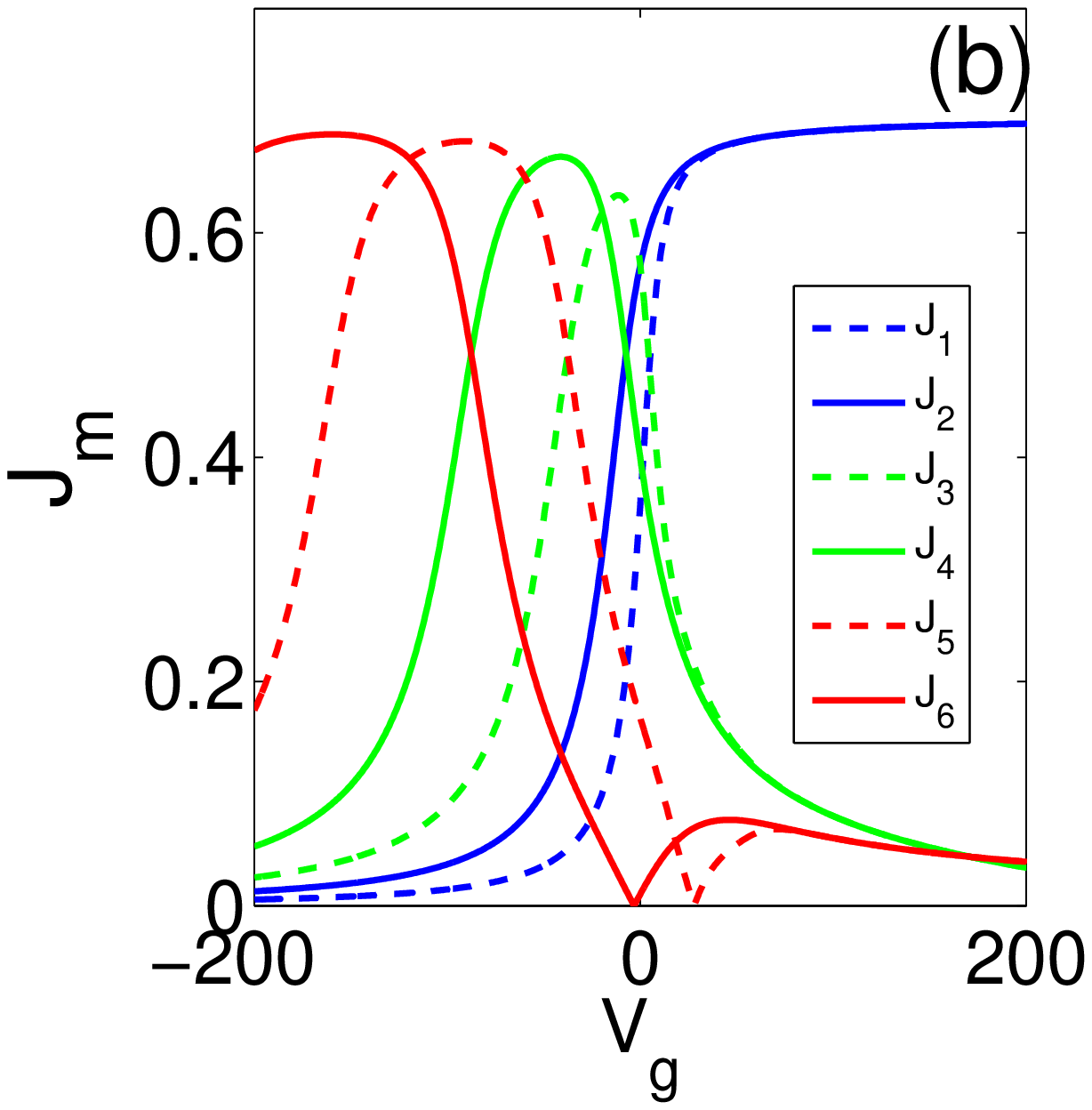}
\caption{(Color online) The first six lowest energy levels (a) and overlapping
integrals (\ref{J}) (b) vs $V_g$. Insert shows
wave function $\phi_5(x)$ for $V_g=0, \pm 200$.}
\label{fig7}
\end{figure}
\begin{equation}\label{J}
J_m=\int_0^1dx \psi_1(x)\phi_m(x)=\sqrt{2}\int_0^1dx \sin\pi x\phi_m(x)
\end{equation}
in the coupling matrix elements in Eq. (\ref{Wmn}) behave non monotonically
as shown in Fig. \ref{fig6} (b). Respectively one can see unusual behavior of the resonances
shown in inset of Fig. \ref{fig5} (a).  As shown in Fig. \ref{fig5} (b) further growth of $V_g$ gives rise to a
hierarchical trapping of resonances predicted by Rotter and co-authors \cite{Iskra,Muller} as a result of the repulsion of complex
eigenvalues of the effective Hamiltonian with the growth of the eigenvalue density. That phenomenon  also takes place
in $\Pi$-shaped waveguides at a different length $L$.

\begin{figure}
\includegraphics[height=5cm,width=5cm,clip=]{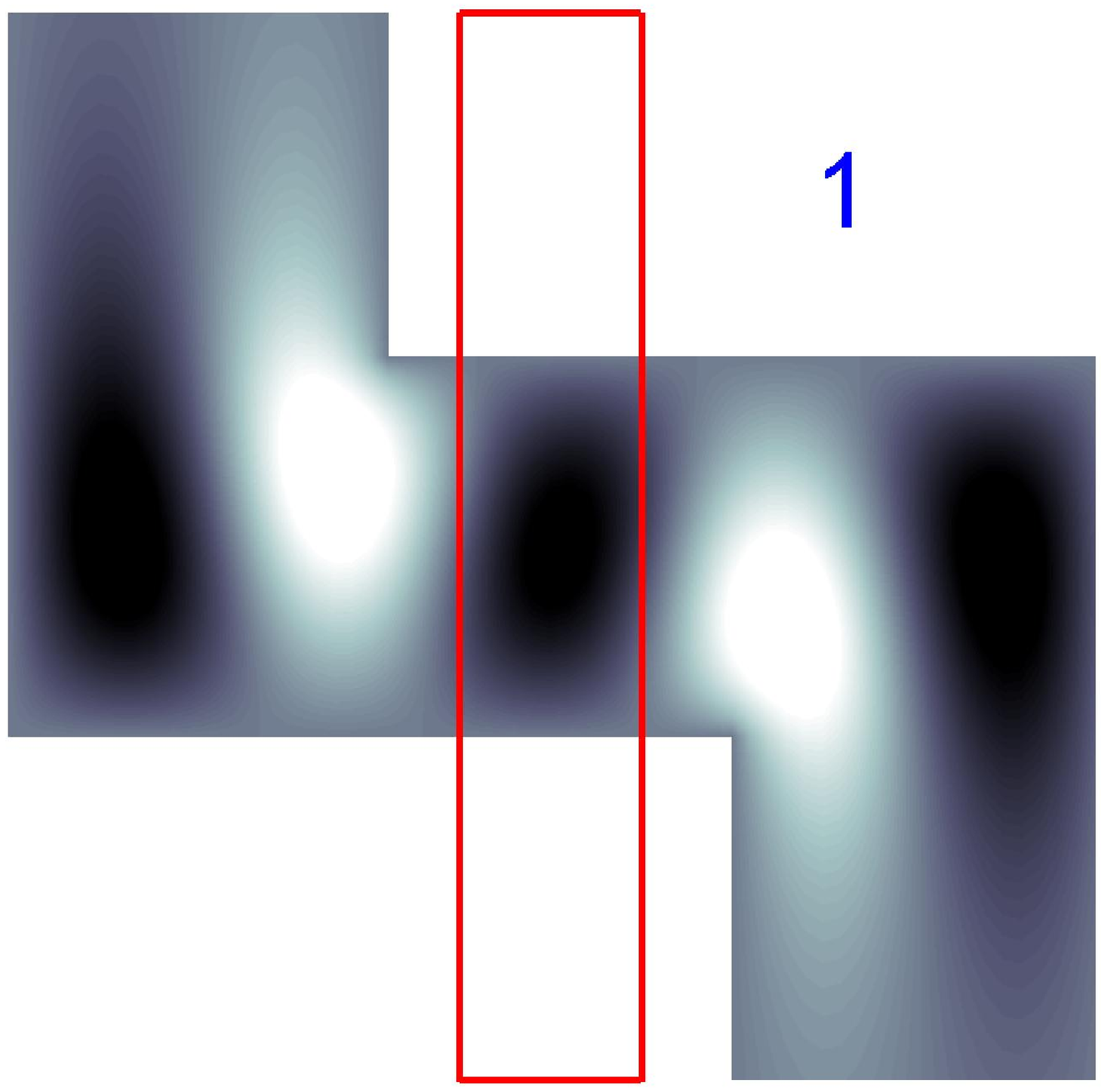}
\includegraphics[height=5cm,width=5cm,clip=]{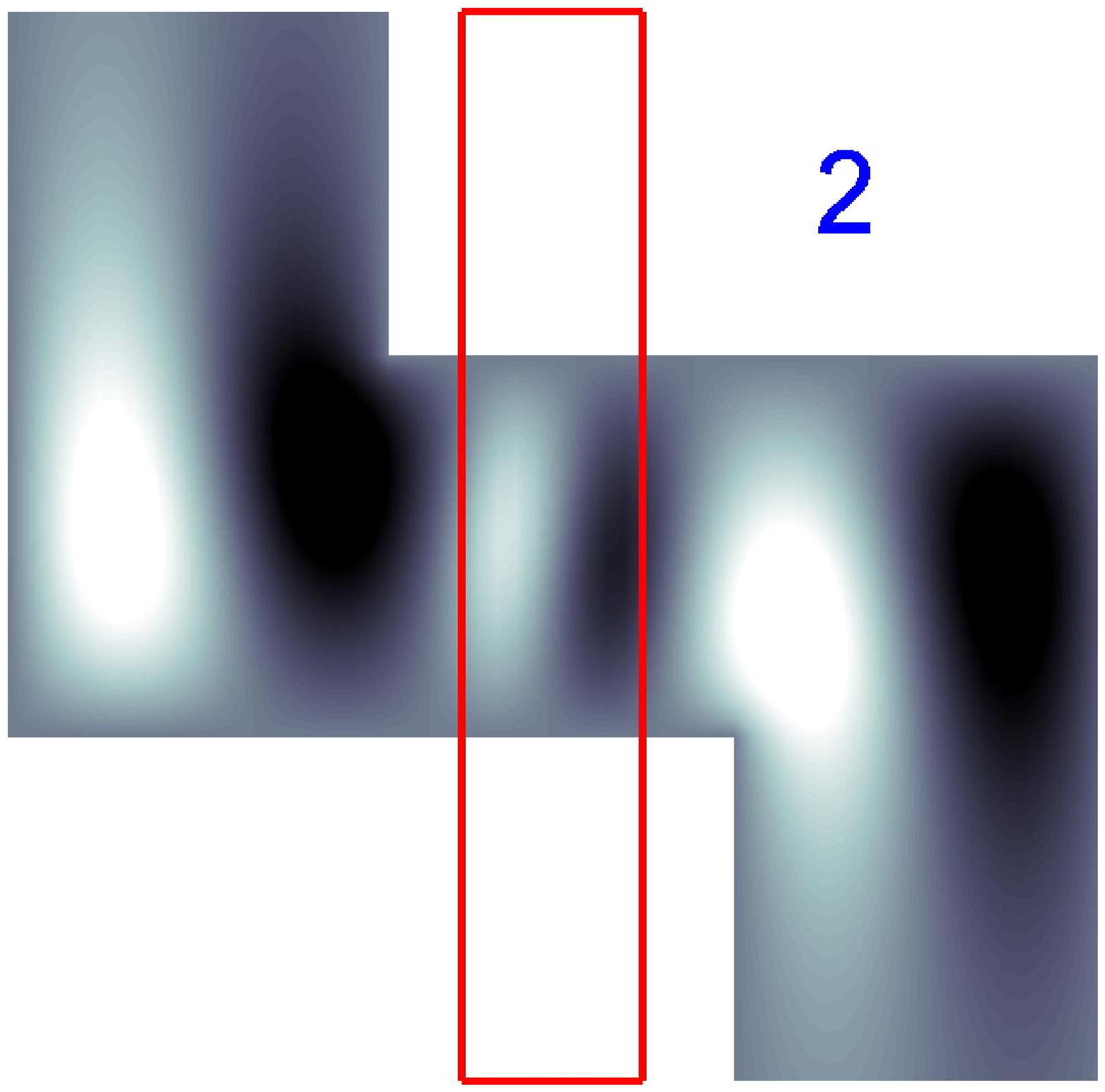}
\includegraphics[height=5cm,width=5cm,clip=]{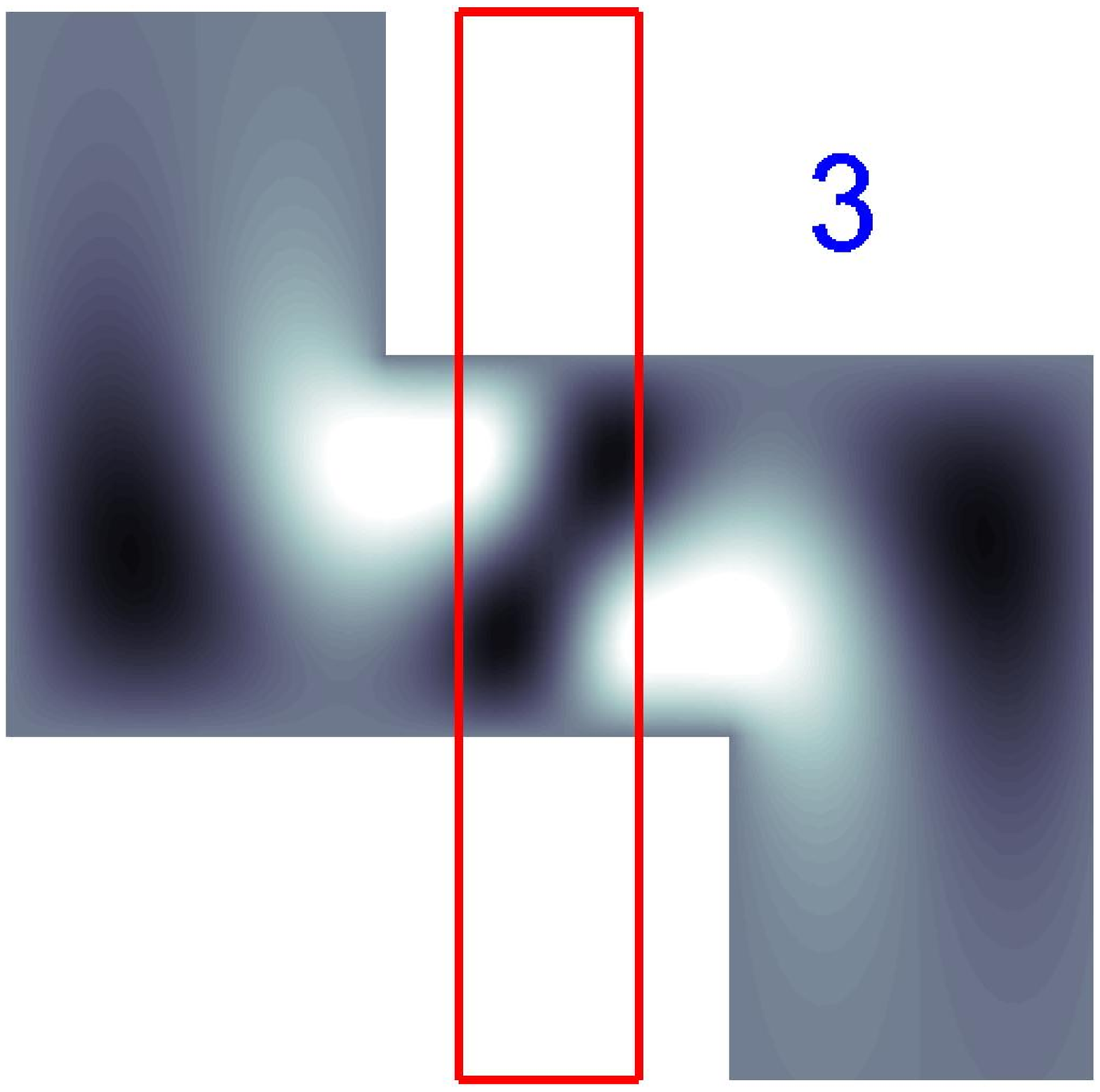}
\includegraphics[height=5cm,width=5cm,clip=]{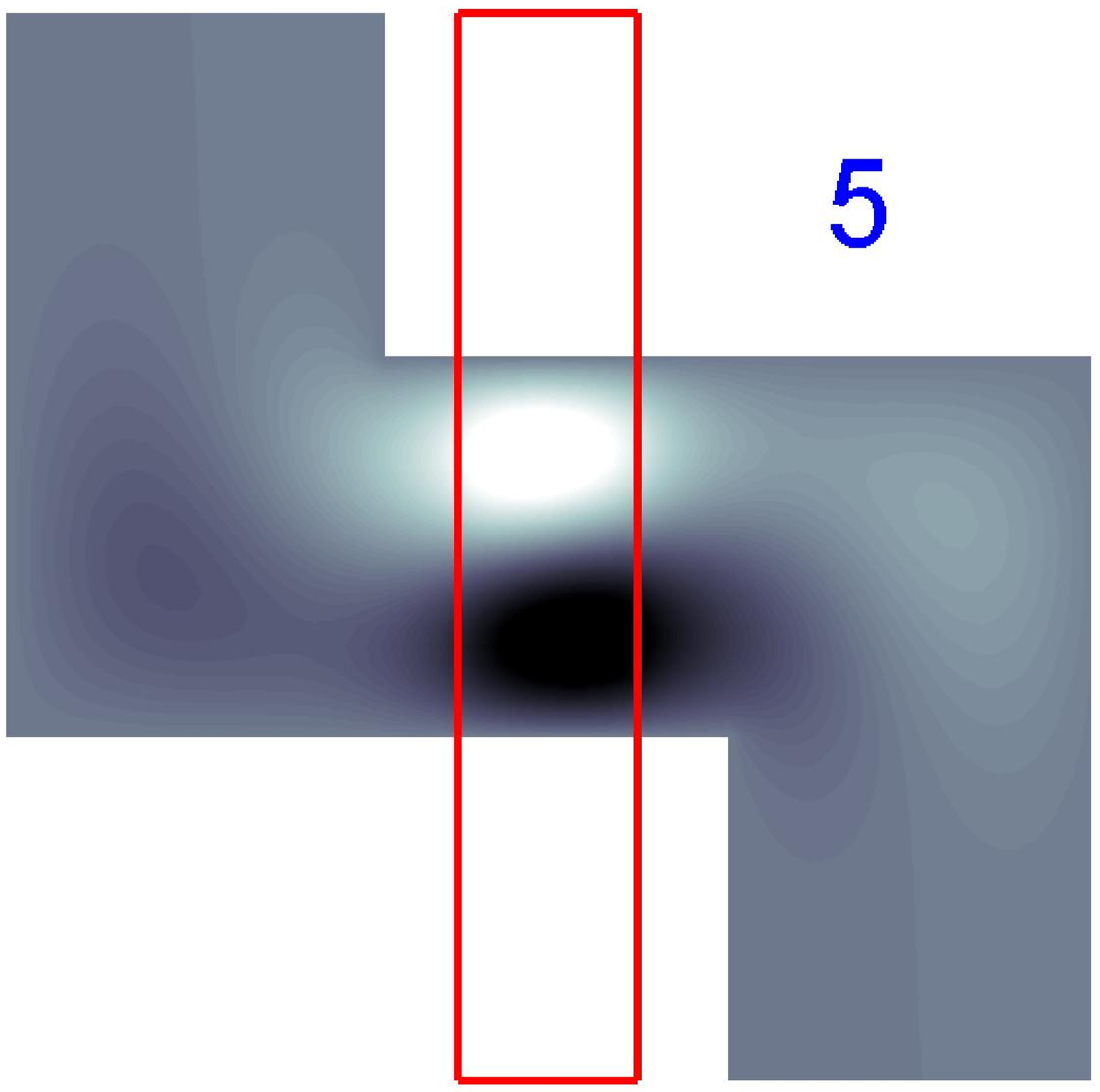}
\includegraphics[height=5cm,width=5cm,clip=]{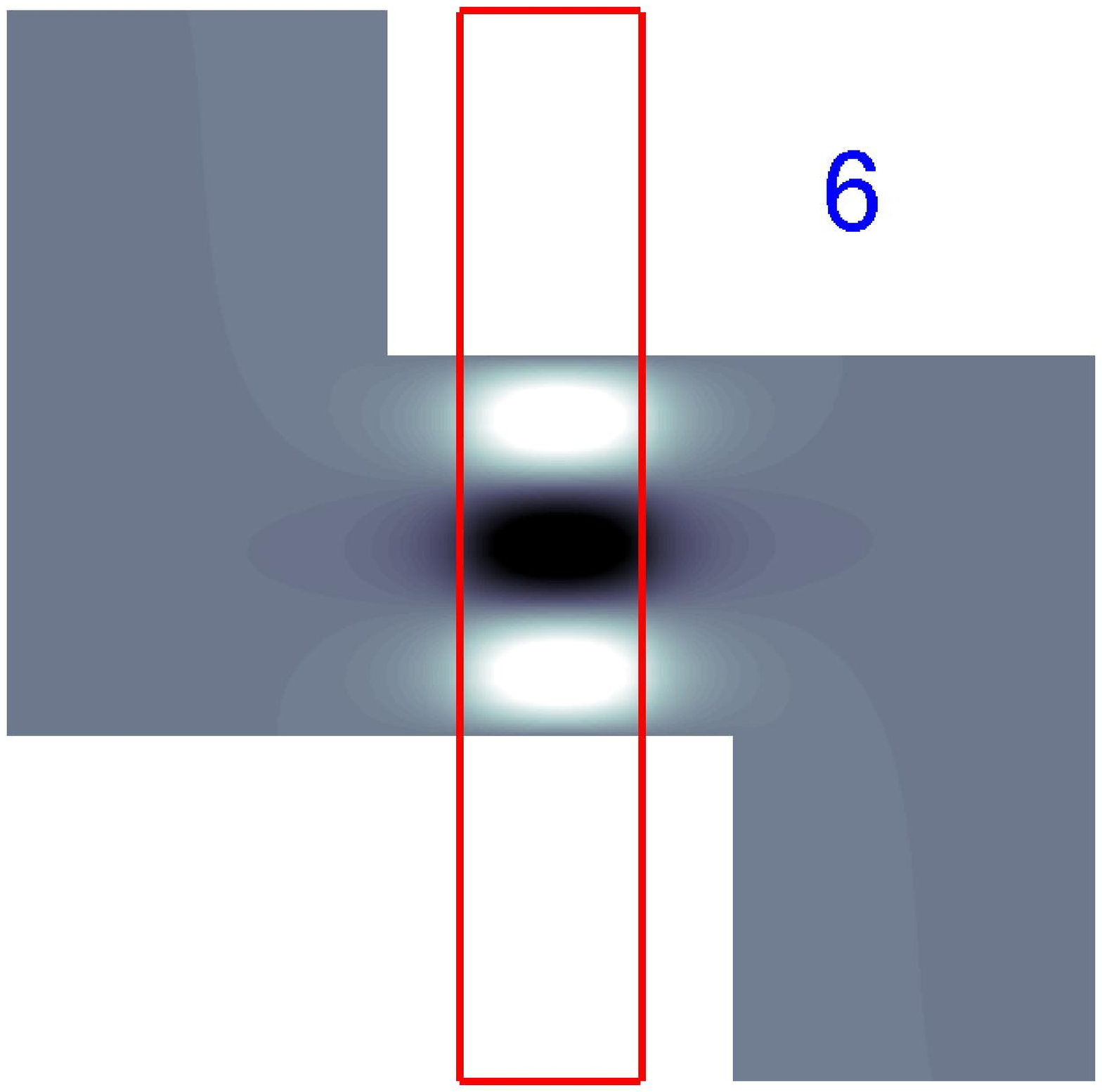}
\includegraphics[height=5cm,width=5cm,clip=]{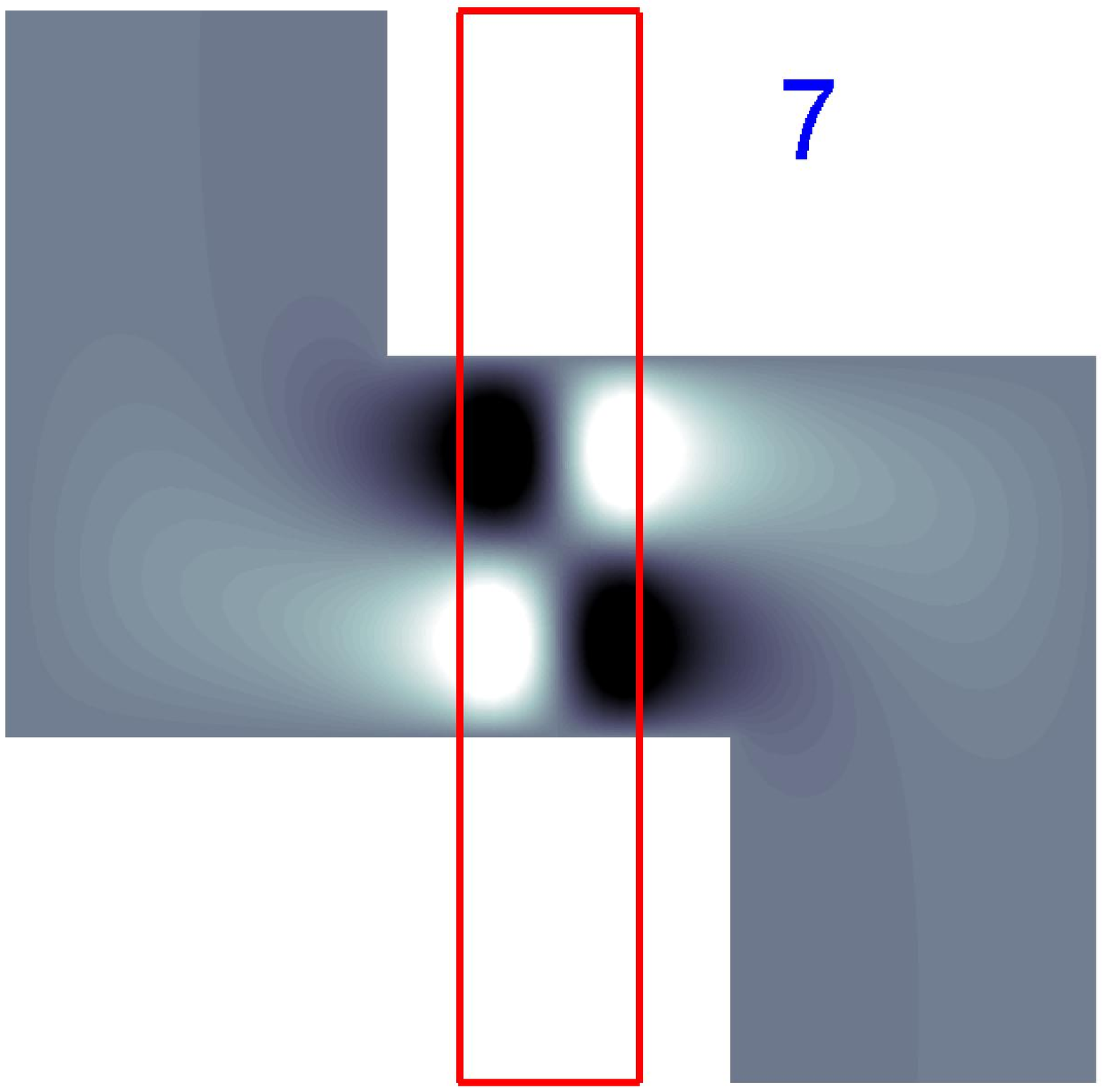}
\caption{(Color online) BSC functions for the $Z$-shaped waveguide with $L=3$ labelled in Fig. \ref{fig7} (a).
 (1) $E=35.066, V_g=-4.8$, (2) $E=35.845, V_g=-83.11$, (3) $E=30.8864, V_g=-68.205$,  (5) $E=24.15, V_g=-26.98$,
(6) $E=16.49, V_g=-90.98$, and (7) $E=16.023, V_g=-97.37$.}
\label{fig8}
\end{figure}

Next, under variation of the the finger gate potential
a unique case can occur when the inverse of matrix $\widehat{H}_{eff}-E$
in Eqs. (\ref{T}) does not exist, when the
determinant $||H_{eff}-E||=0$. That corresponds to $Im[z(E_{BSC})]=0$ and $E_{BSC}=Re[z(E_{BSC})]$.
These equations define the necessary and sufficient condition
for the  BSC \cite{ring,PRA} and can be combined into a single equation for the BSC point
\begin{equation}\label{abs}
    |H_{eff}(E_{BSC})-E_{BSC}|=0.
\end{equation}
The corresponding eigenfunction of the effective Hamiltonian
is the BSC function, respectively. The solution
Fig. \ref{fig6} (c) demonstrates multiple BSCs denoted by red open circles occurring under variation of $V_g$.
An alternative approach to
diagnose the BSC is calculating the poles of the S-matrix and finding the condition under which a complex pole
tends to the real axis \cite{Pursey,Ordonez,Cattapan,hatano,Lee&Reichl,Hein}.
Clearly, as the imaginary part of a pole
or a complex eigenvalue tends to zero both methods yield identical results.
The BSCs can be also be found in the conductance vs. the Fermi energy and the potential $V_g$ as the singular points where
full reflection ($G=0$) meets the full transmission ($G=1$) \cite{PRB,Lepetit,Lepetit1}. These
points are singular due to the collapse of the Fano resonances \cite{kim1}.
In Fig. \ref{fig7} (a) these points are marked by white circles and enumerated to show
corresponding BSC functions in Fig. \ref{fig8}. The BSC function 3 is very similar to the
BSC function 4 and is not shown in Fig. \ref{fig8}.

In Fig. \ref{fig8} one can see two different types of BSCs. The first three 1-3 are due to the FPR mechanism discussed
in the Introduction. The other three BSC functions 5-7 are reminiscent to those considered by Robnik \cite{Robnik} in a directional quantum
waveguide with a negative potential $V_g$  shown in Fig. \ref{fig1} (c). In this case
the Schr\"odinger equation is separable
\begin{equation}\label{rob1}
    \psi(x,y)=\sin(\pi ny/d)\phi(x)
\end{equation}
which gives rise to the effectively one-dimensional transmission through a potential well
\begin{equation}\label{rob2}
\frac{\partial^2\phi}{\partial x^2}+[\epsilon-V_g(x)]\psi=0
\end{equation}
where
\begin{equation}\label{rob3}
\epsilon=E-\frac{\pi^2 n^2}{d^2}.
\end{equation}
There are approximately $N=L\sqrt{-V_g}/\pi$ one-dimensional bound states $\phi_m(x)$ with discrete
energies $\epsilon_m$ in the potential well with length $L_g$ in terms of the width of the waveguide $d$
and depth $V_g$ for $V_g<0$ \cite{Robnik}. All bound states with $n=1$ and $\epsilon_m<0$
are  below the first channel propagation band and therefore are not BSCs. But
the bound states with $n=2,3 \ldots$ and energies $4\pi^2>\pi^2n^2+\epsilon_m$ become
BSC because of their orthogonality to the first channel propagation state
with $n=1$.

\begin{figure}
\includegraphics[height=5cm,width=5cm,clip=]{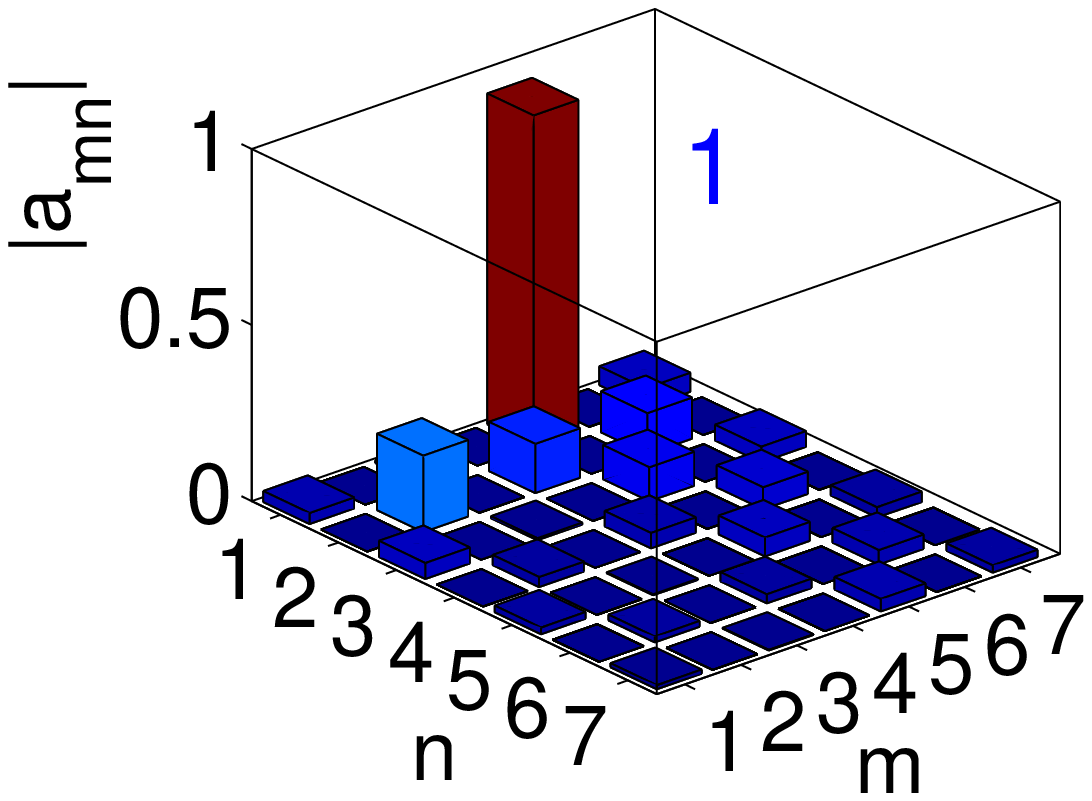}
\includegraphics[height=5cm,width=5cm,clip=]{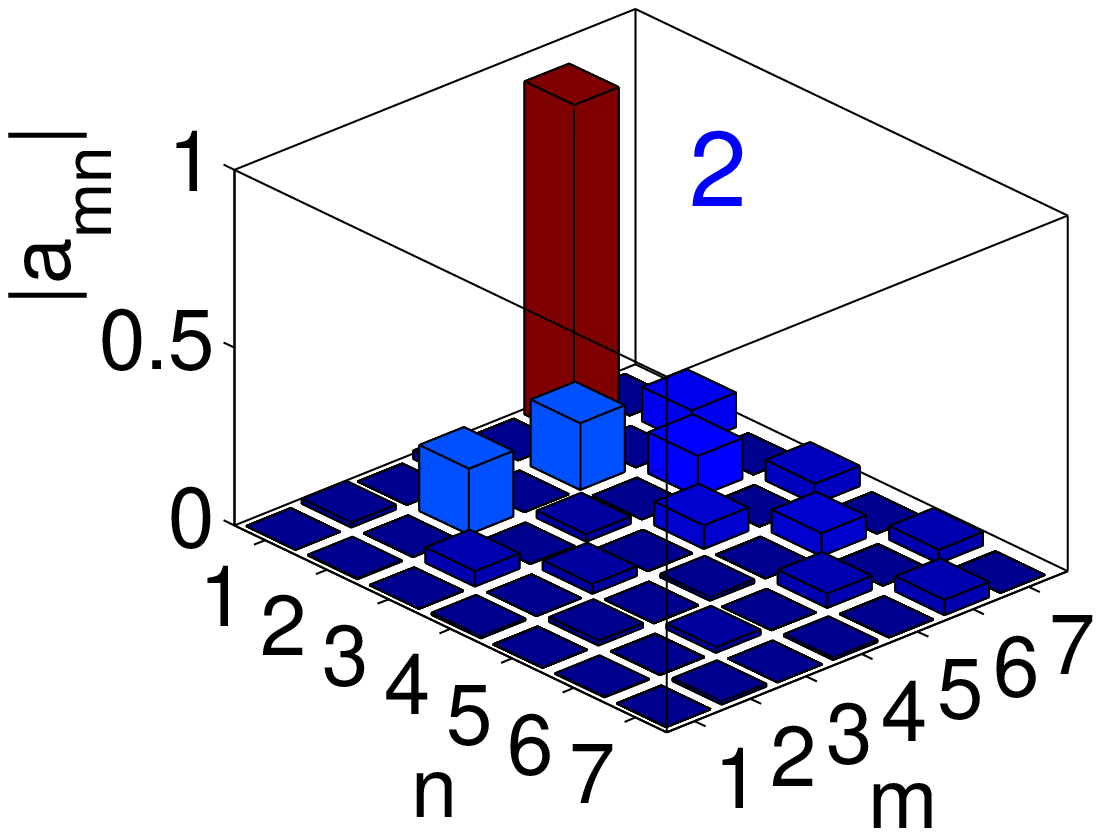}
\includegraphics[height=5cm,width=5cm,clip=]{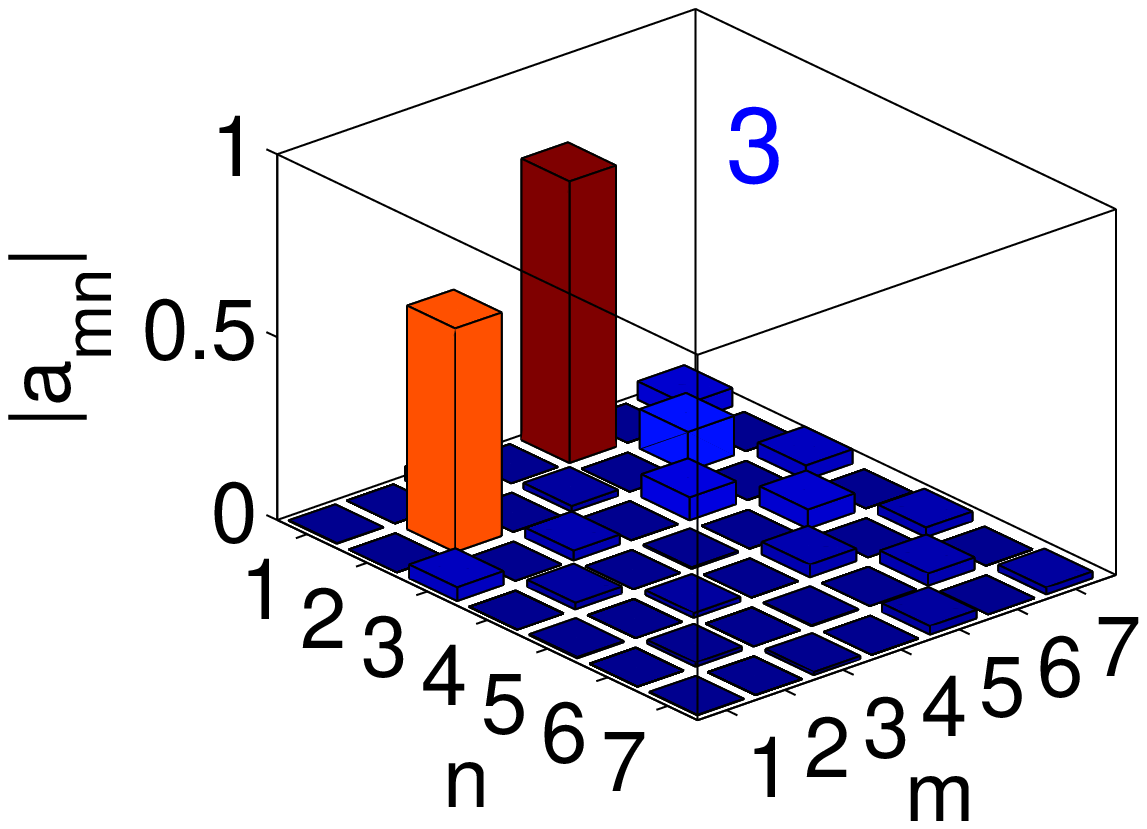}
\includegraphics[height=5cm,width=5cm,clip=]{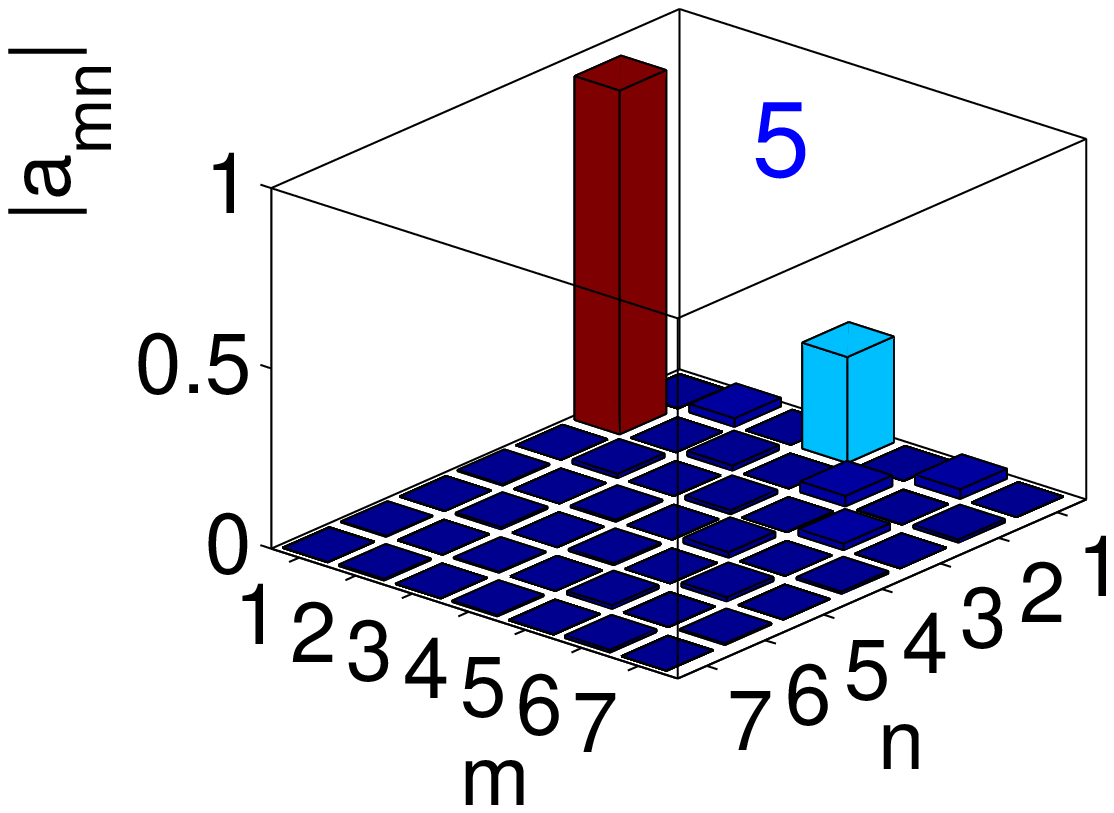}
\includegraphics[height=5cm,width=5cm,clip=]{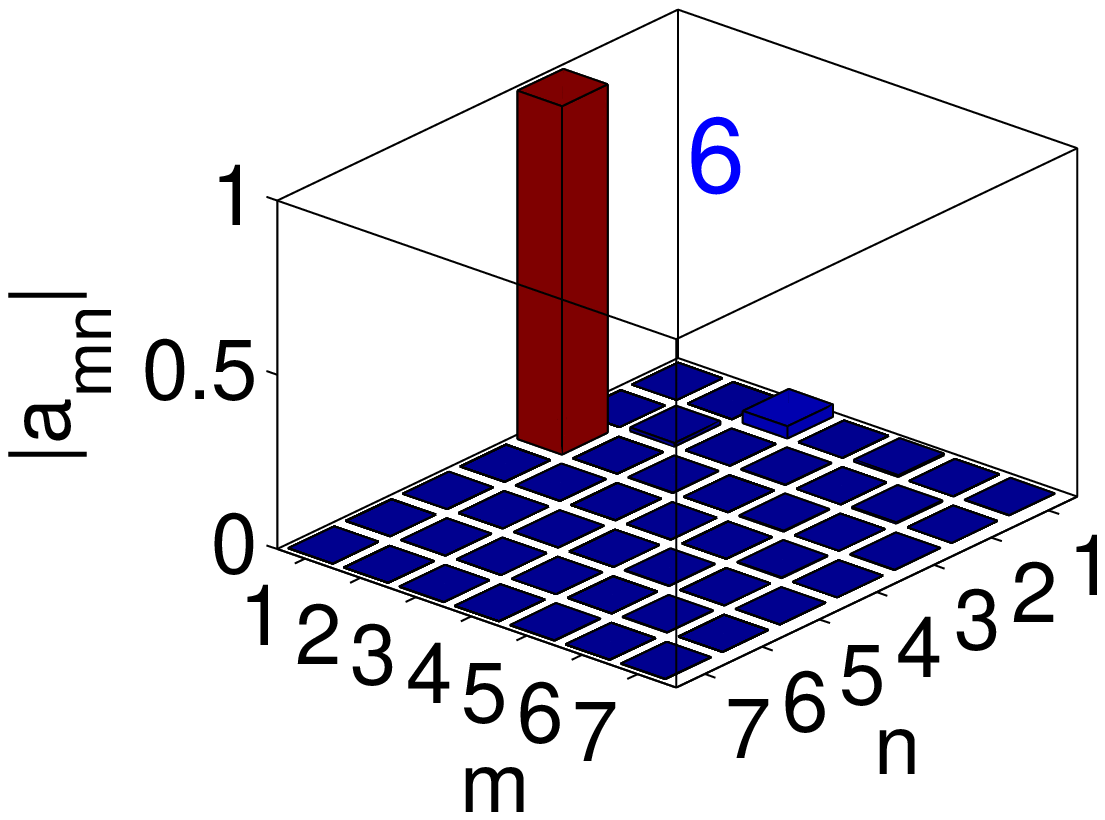}
\includegraphics[height=5cm,width=5cm,clip=]{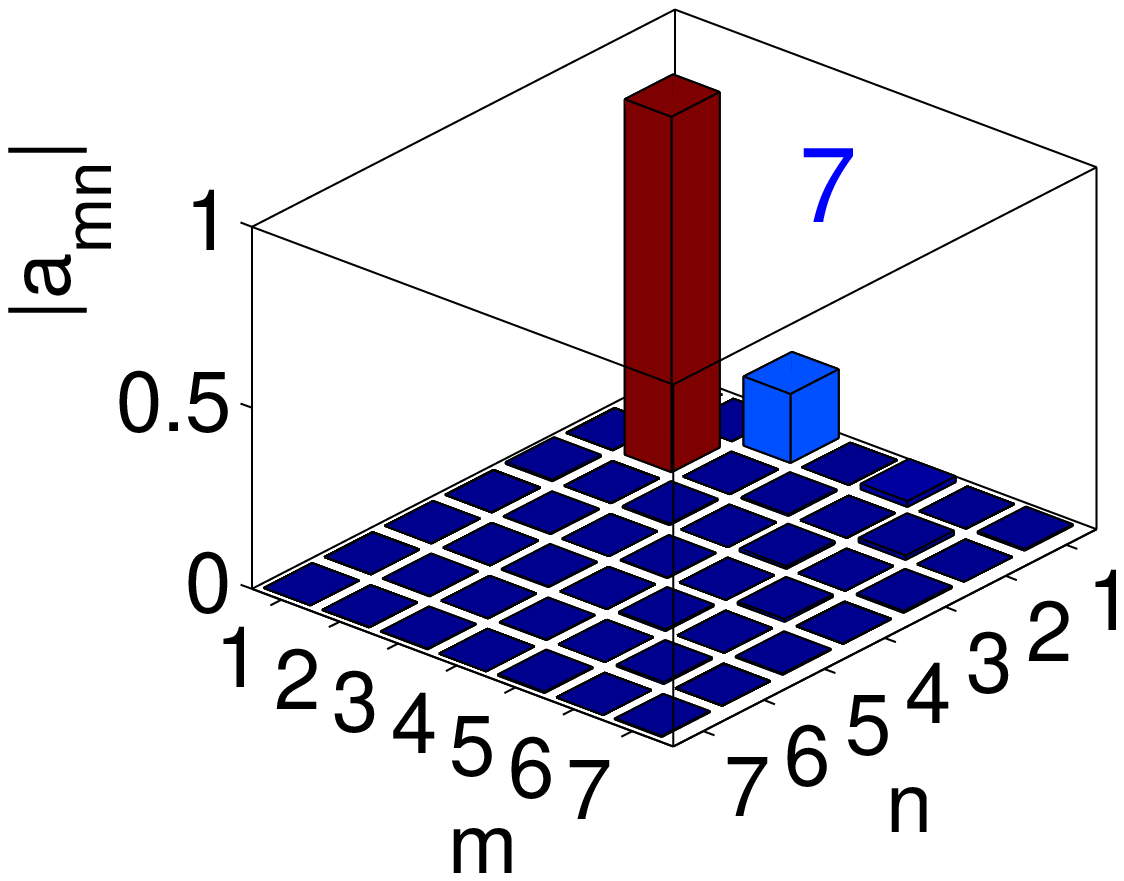}
\caption{(Color online) Expansion of BSC functions shown in Fig. \ref{fig8} over the
bridge eigenfunctions (\ref{psimn}).}
\label{fig9}
\end{figure}

Turning back to the double-bent waveguides the symmetry arguments by Robnik are not valid
because the coupling of some bridge state with the first channel can be cancelled only accidentally.
The numerics show that it only occurs for the Fermi
energy above the second channel threshold $4\pi^2$ as shown in Fig. \ref{fig7} (a). The reason is clear and
can be readily seen from Eq. (\ref{Wmn}). For the coupling matrix element (\ref{Wmn}) to be equal to zero
it is necessary that the wave function of the bridge $\phi_m(x)$ had a nodal point at $x=1/2$ (see inset
Fig. \ref{fig6}). That obviously corresponds to the second channel transmission as
shown in Fig. \ref{fig7} (a) by red closed circles.
Therefore the BSC at $V_g<0$ are realized through the Friedrich-Wintgen
interference mechanism. As  Figs. \ref{fig8} and Fig. \ref{fig9} show the bridge
wave function with longitudinal quantum number $m=1$ dominates in BSC functions 5 and 6. For
the corresponding finger gate potentials given in the figure caption these function are mostly localized
underneath the gate and decay exponentially in the waveguides. Therefore the
contributions of the other bridge states to decouple the resonant state from
the first propagation channel has to be also exponentially small as seen from  Fig. \ref{fig9} for BSCs
5 and 6.

\begin{figure}
\includegraphics[height=5cm,width=5cm,clip=]{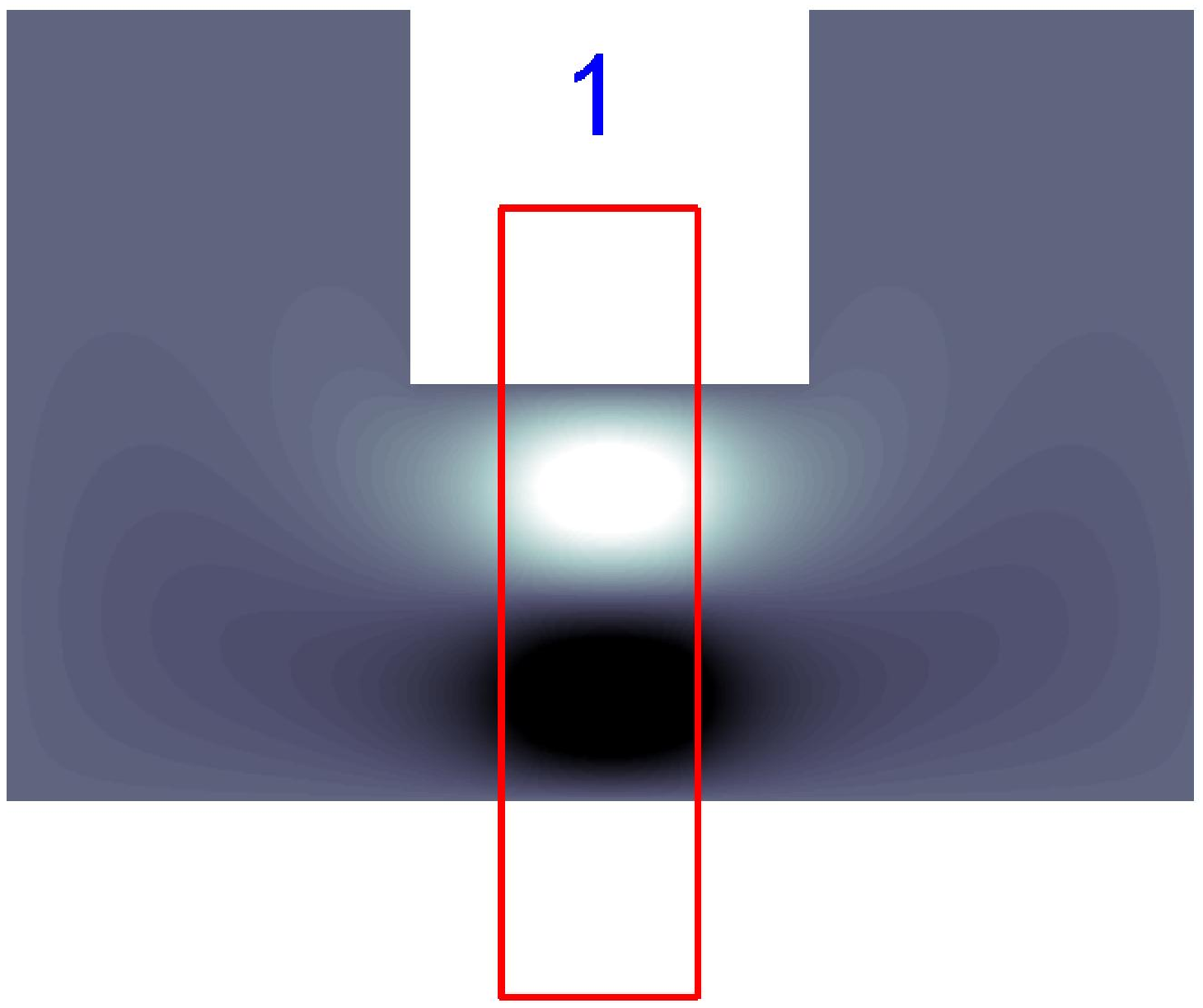}
\includegraphics[height=5cm,width=5cm,clip=]{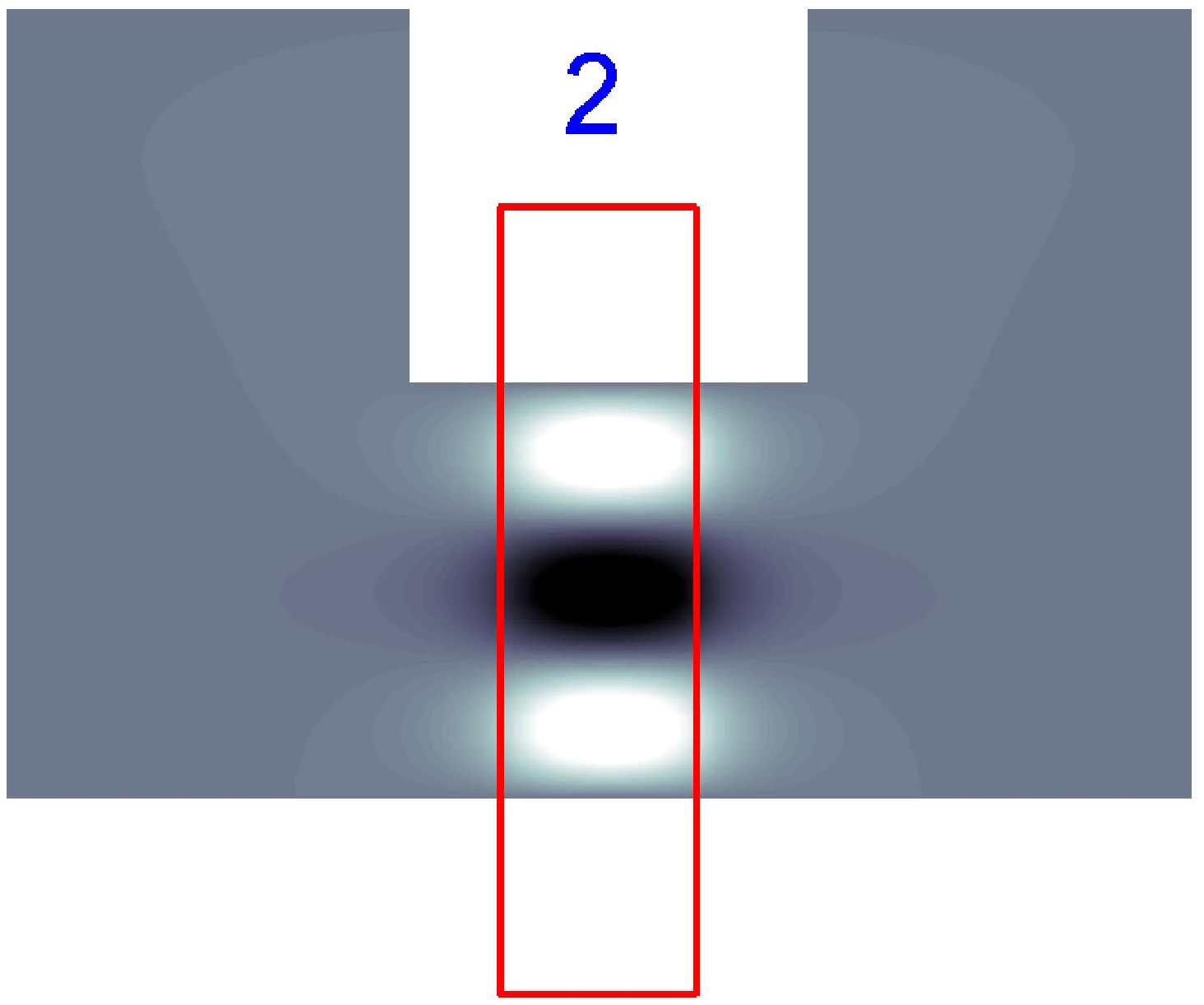}
\includegraphics[height=5cm,width=5cm,clip=]{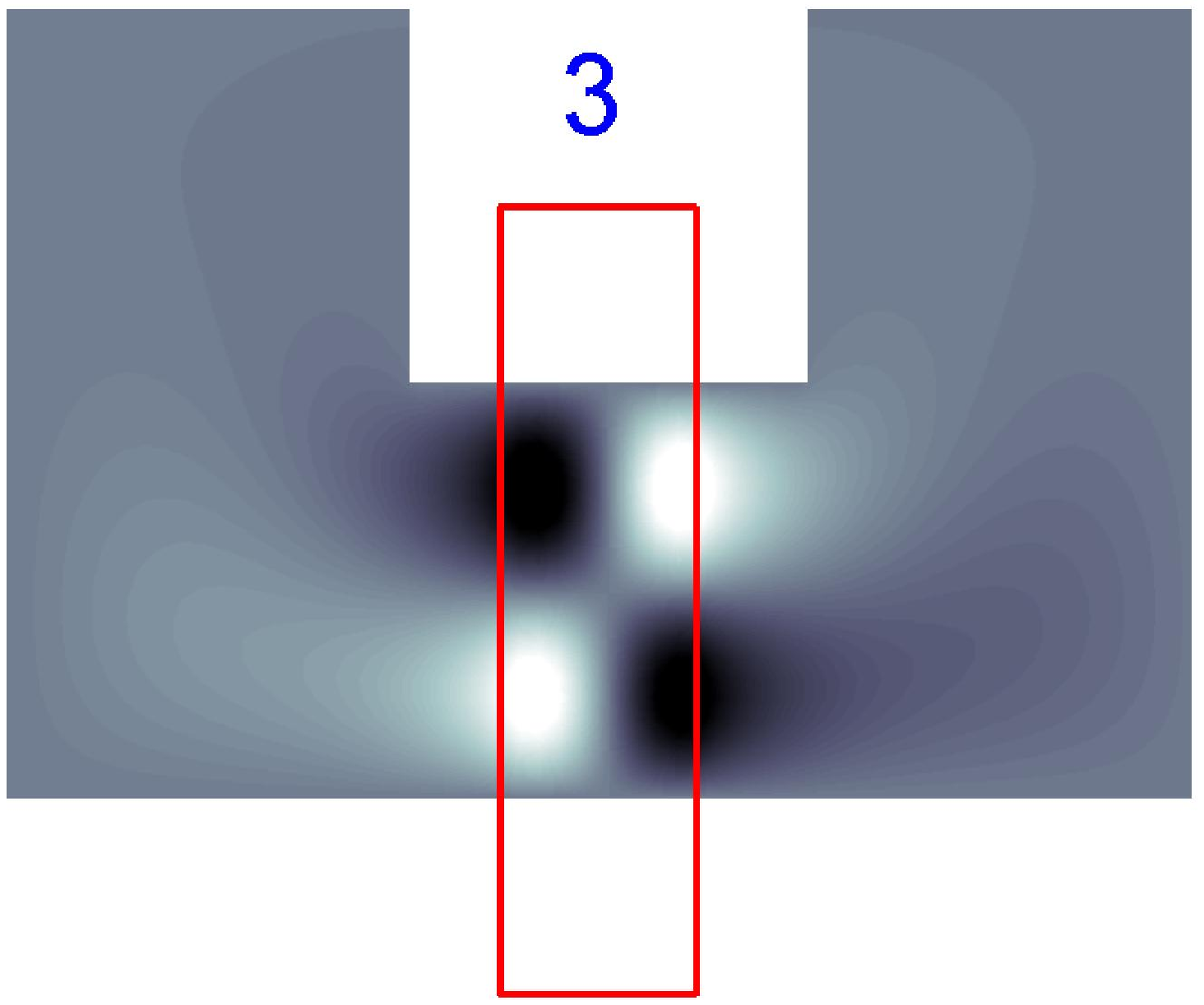}
\caption{(Color online) BSC functions for the $\Pi$-shaped waveguide with $L=3$ labelled in Fig. \ref{fig7} (b).
(1) $E=17.24, V_g=-35.86$,
(2) $E=16.86, V_g=-90.58$, (4) $E=29.81, V_g=-76.49$.}
\label{fig10}
\end{figure}

\begin{figure}
\includegraphics[height=5cm,width=5cm,clip=]{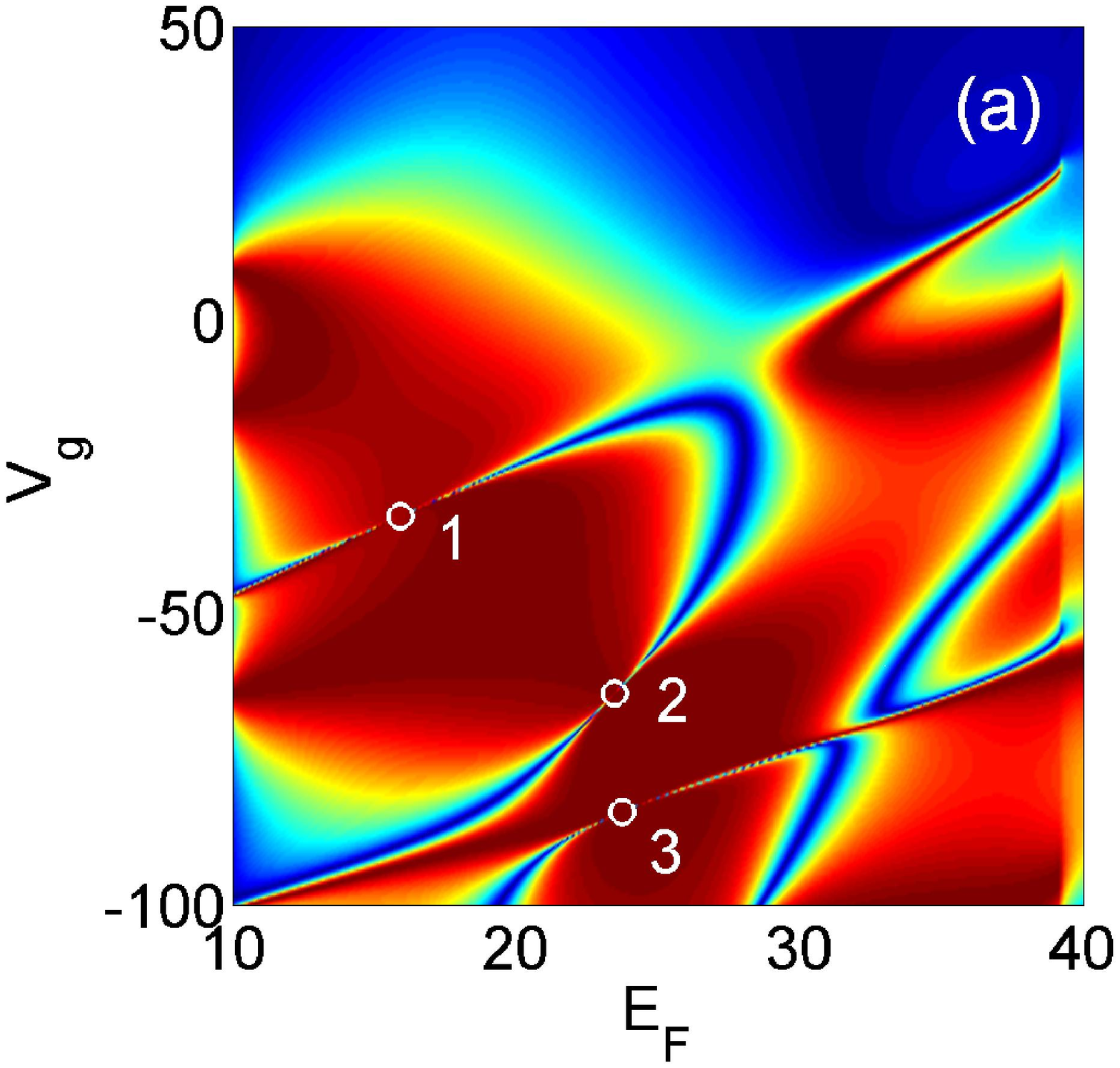}
\includegraphics[height=5cm,width=5cm,clip=]{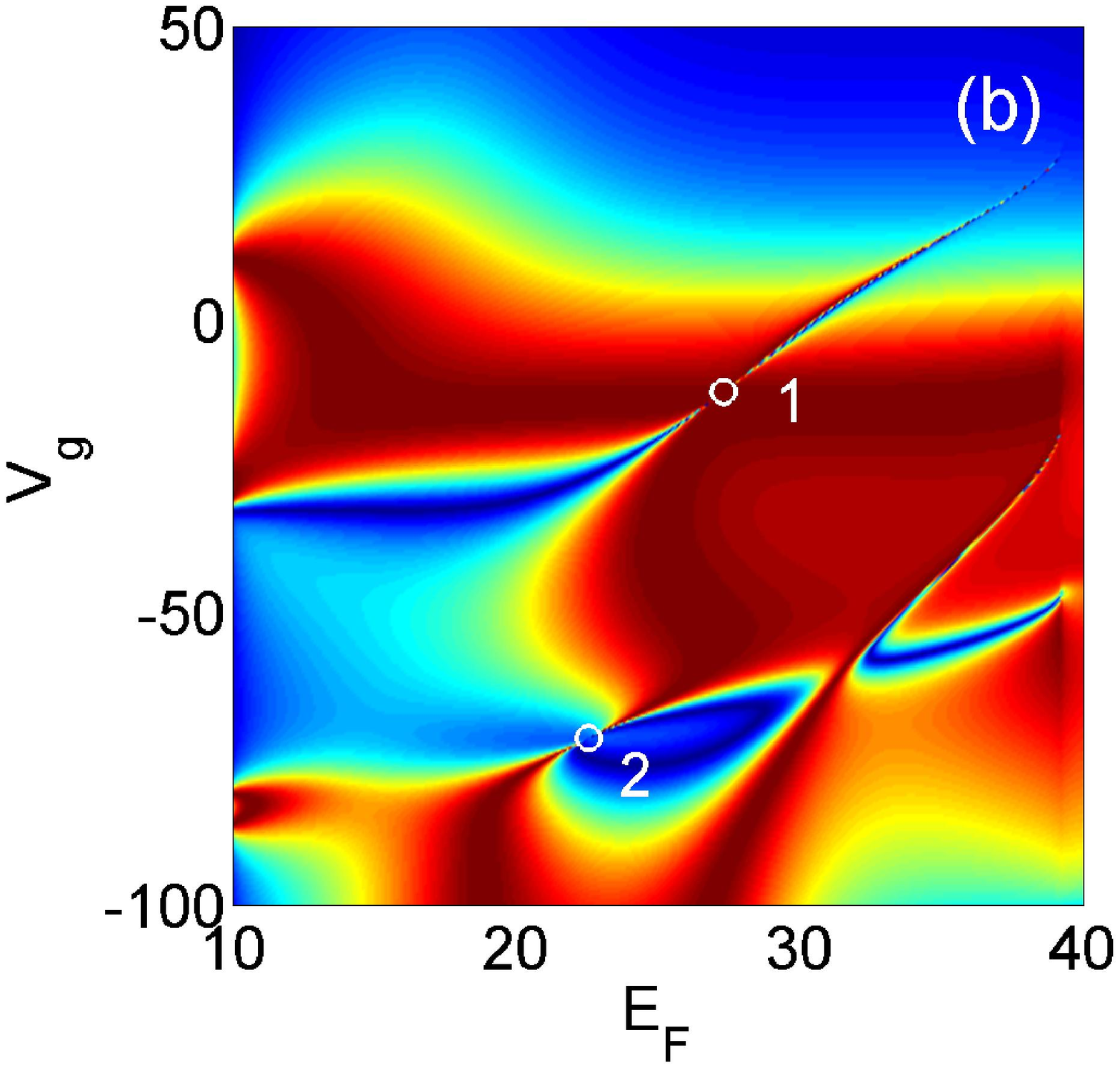}
\caption{(Color online) Conductance of double-bend waveguide vs Fermi energy and $V_g$
for $Z$- (a) and  $\Pi$-shaped waveguides (b) for $L=2$. Open circles mark BSCs. }
\label{fig11}
\end{figure}

\begin{figure}
\includegraphics[height=5cm,width=5cm,clip=]{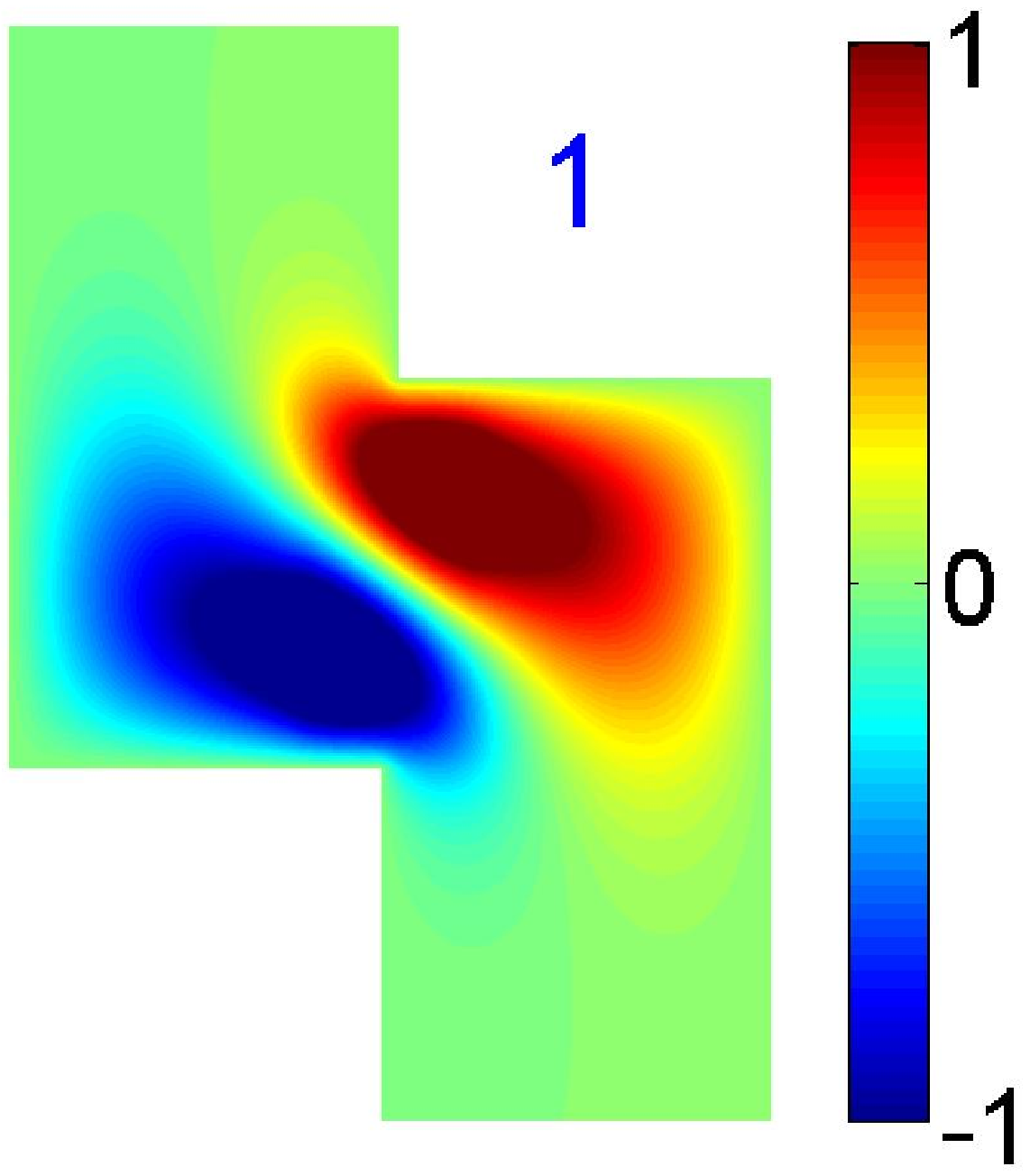}
\includegraphics[height=5cm,width=5cm,clip=]{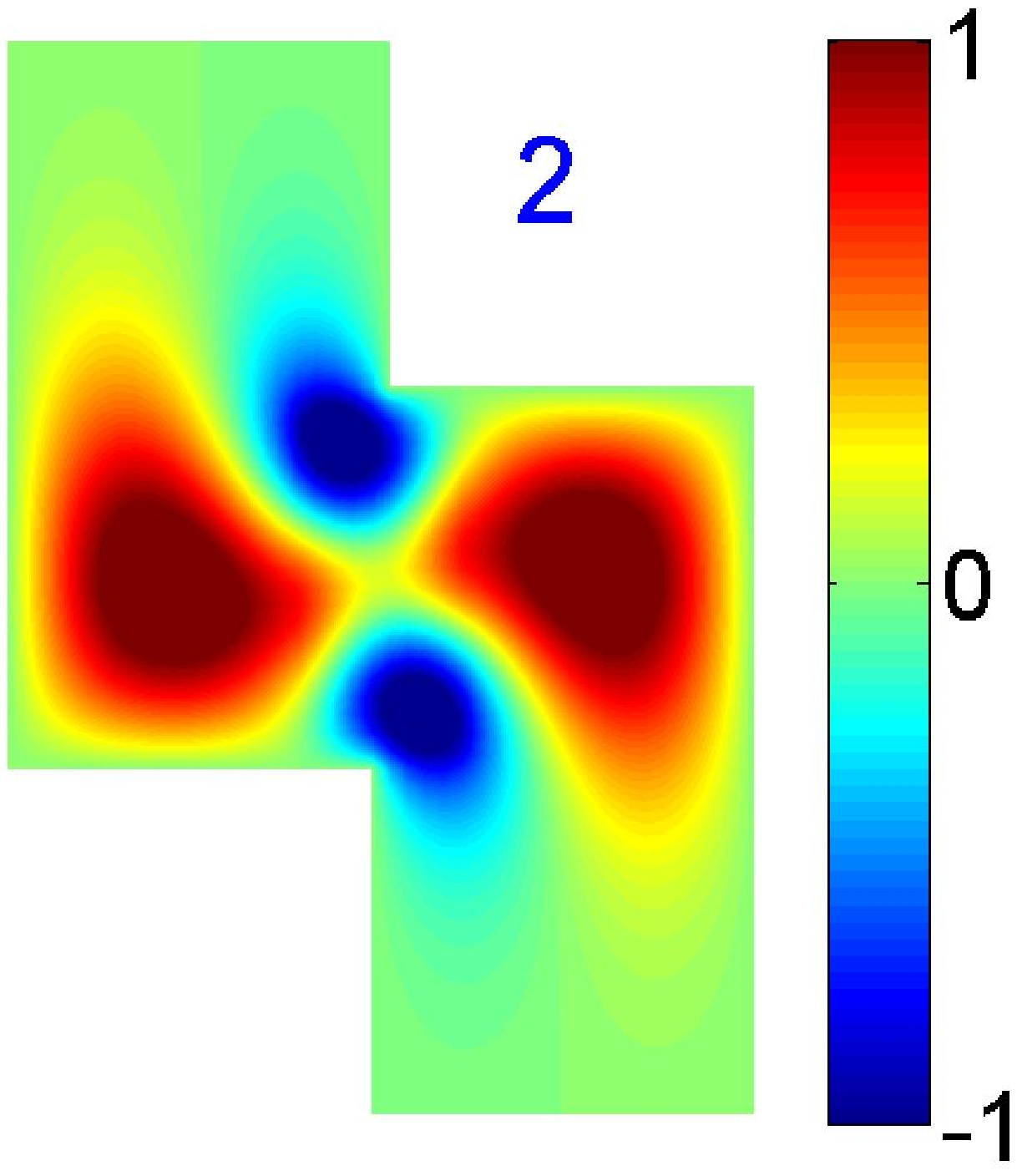}
\includegraphics[height=5cm,width=5cm,clip=]{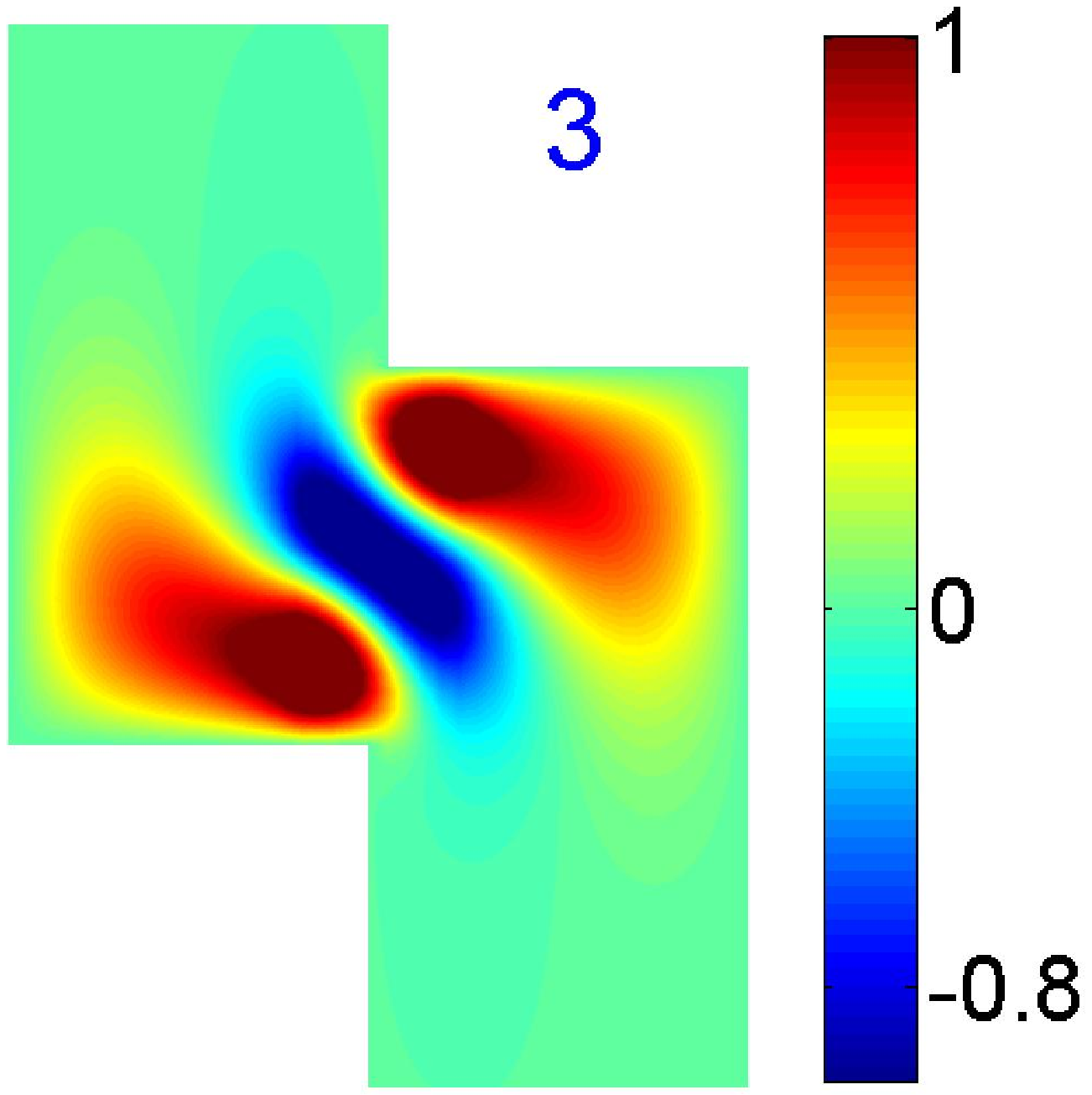}
\caption{(Color online) BSC functions for the $Z$-shaped waveguide labelled in Fig. \ref{fig11} (a).
 (1) $E=15.9213, V_g=-33.54$,
(2) $E=23.464, V_g=-63.75$, (3) $E=23.73, V_g=-83.932$.}
\label{fig12}
\end{figure}

The BSC functions in the $\Pi$-shaped waveguide are shown in Fig. \ref{fig10}.
For $L=3$ and the chosen range of the finger gate potential the $\Pi$-shaped waveguide
does not display the FPR type of the BSCs. Nevertheless the FPR type BSCs occur for $L=5$.
Finally in Fig. \ref{fig11} we show the conductance of the shortest double-bend waveguides
$L=2$ with the corresponding BSCs in Fig. \ref{fig12} for the $Z$-shaped
waveguide and Fig. \ref{fig14} for the $\Pi$-shaped waveguide, respectively.
\begin{figure}
\includegraphics[height=5cm,width=5cm,clip=]{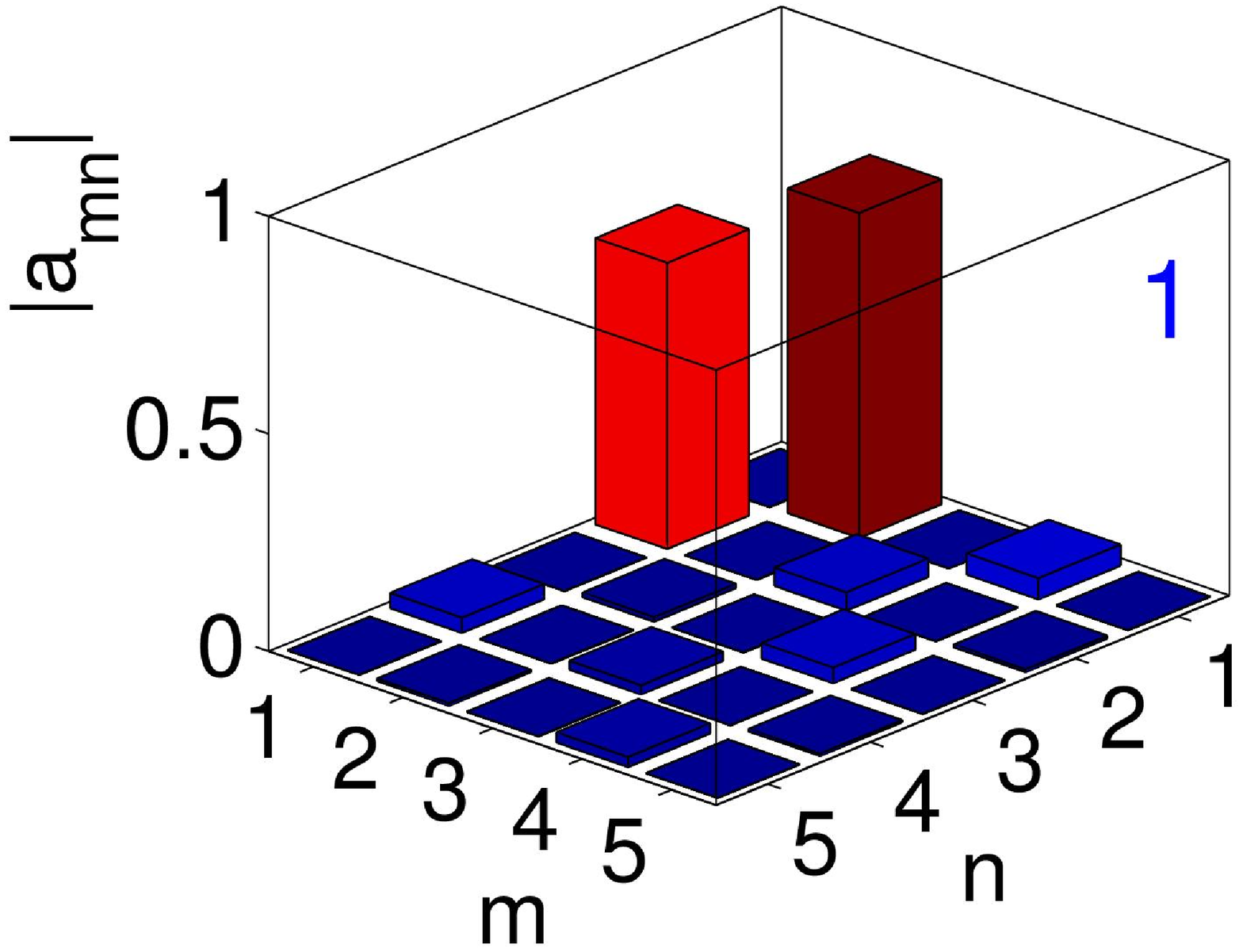}
\includegraphics[height=5cm,width=5cm,clip=]{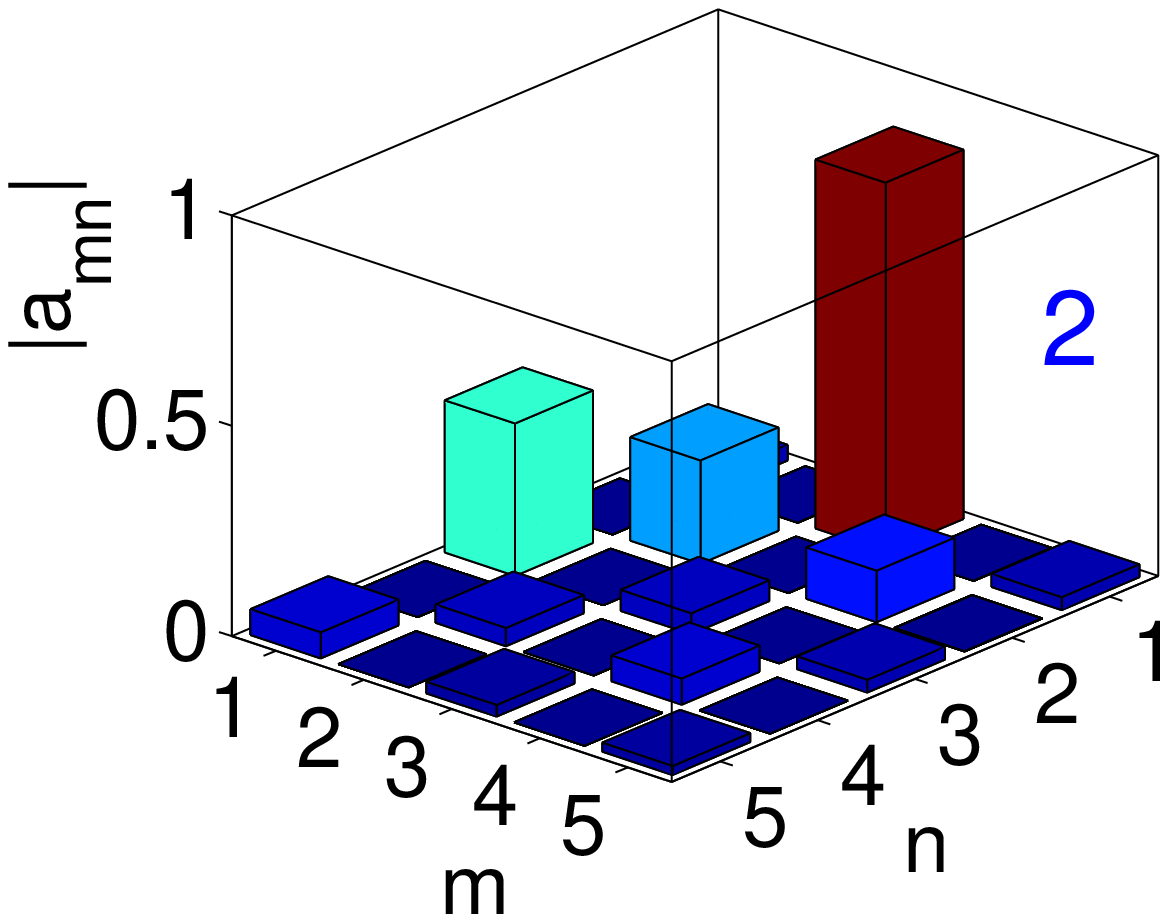}
\includegraphics[height=5cm,width=5cm,clip=]{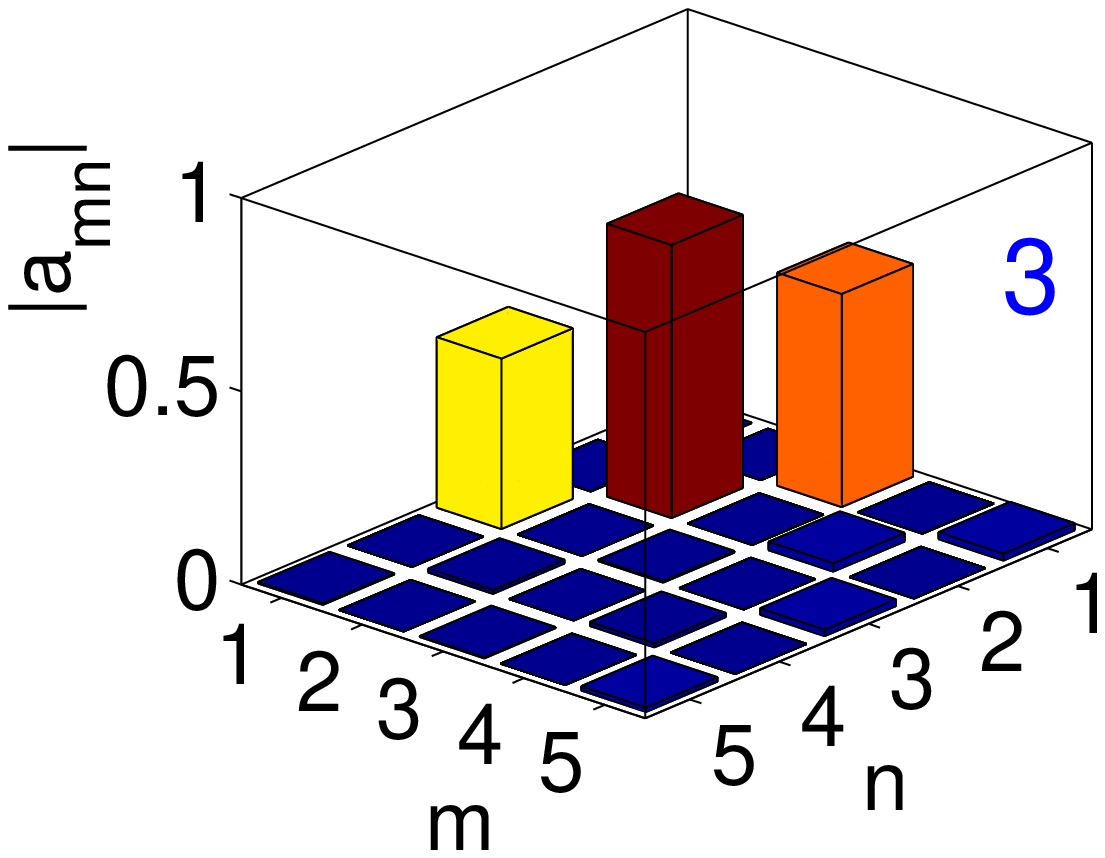}
\caption{(Color online) Expansion of BSC functions in $Z$-shaped
waveguide shown in Fig. \ref{fig13} over the inner eigenfunctions (\ref{psimn}).}
\label{fig13}
\end{figure}
\begin{figure}
\includegraphics[height=6cm,width=6cm,clip=]{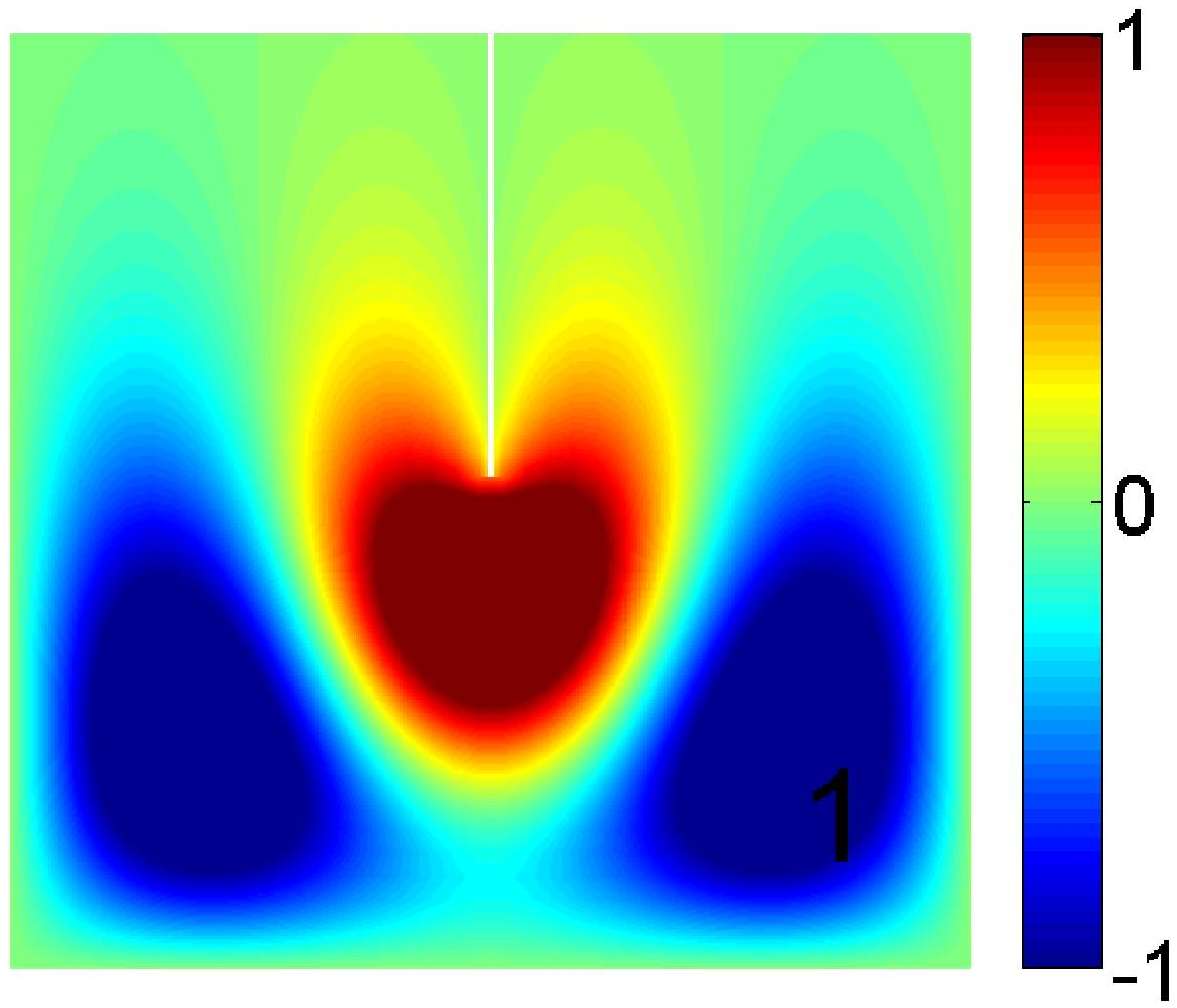}
\includegraphics[height=6cm,width=6cm,clip=]{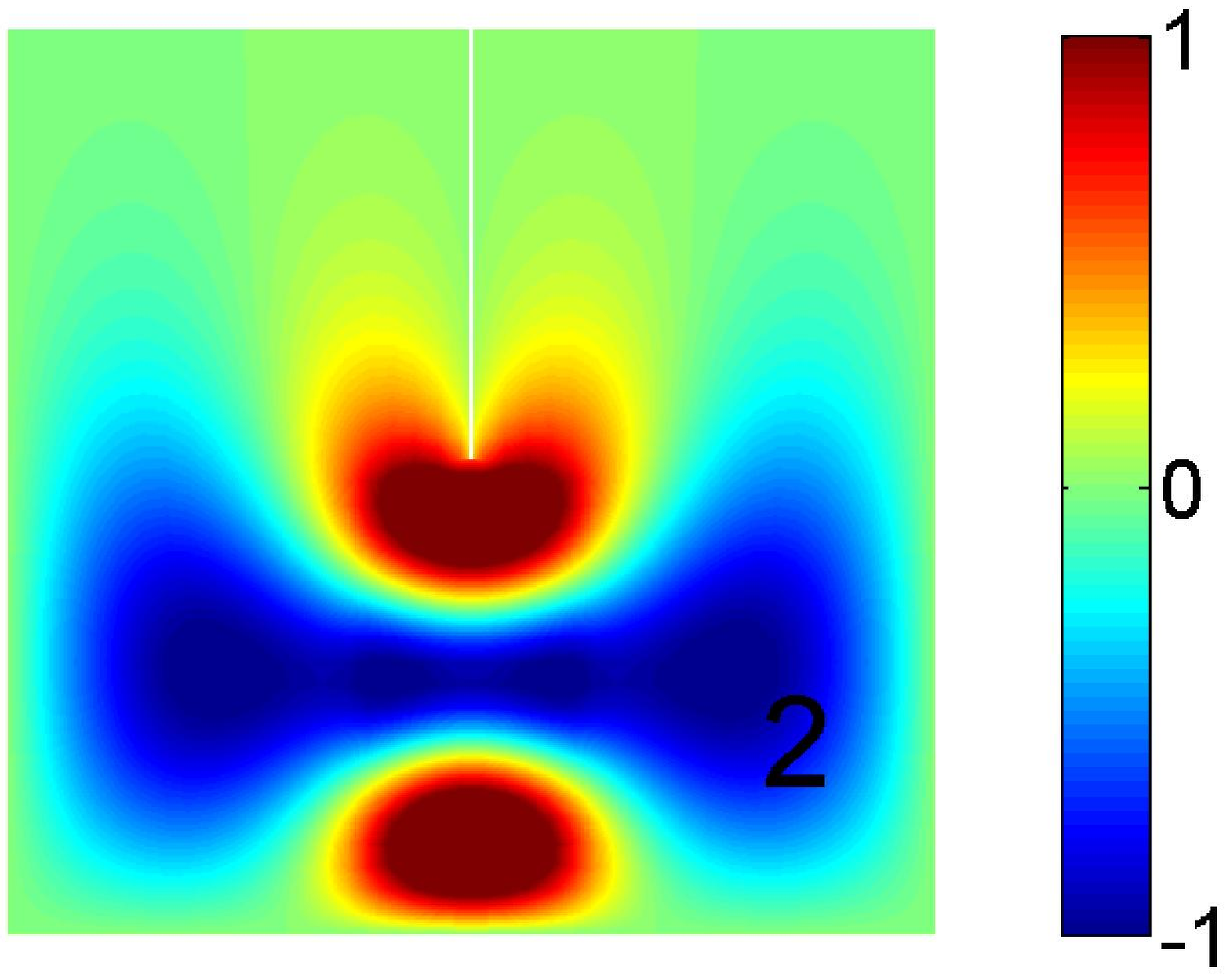}
\caption{(Color online) BSC functions for the $\Pi$-shaped waveguide labelled in Fig. \ref{fig11} (b).
 (1) $E=27.3, V_g=-12.31$, (2) $E=22.55, V_g=-71.34$.}
\label{fig14}
\end{figure}
\begin{figure}
\includegraphics[height=5cm,width=5cm,clip=]{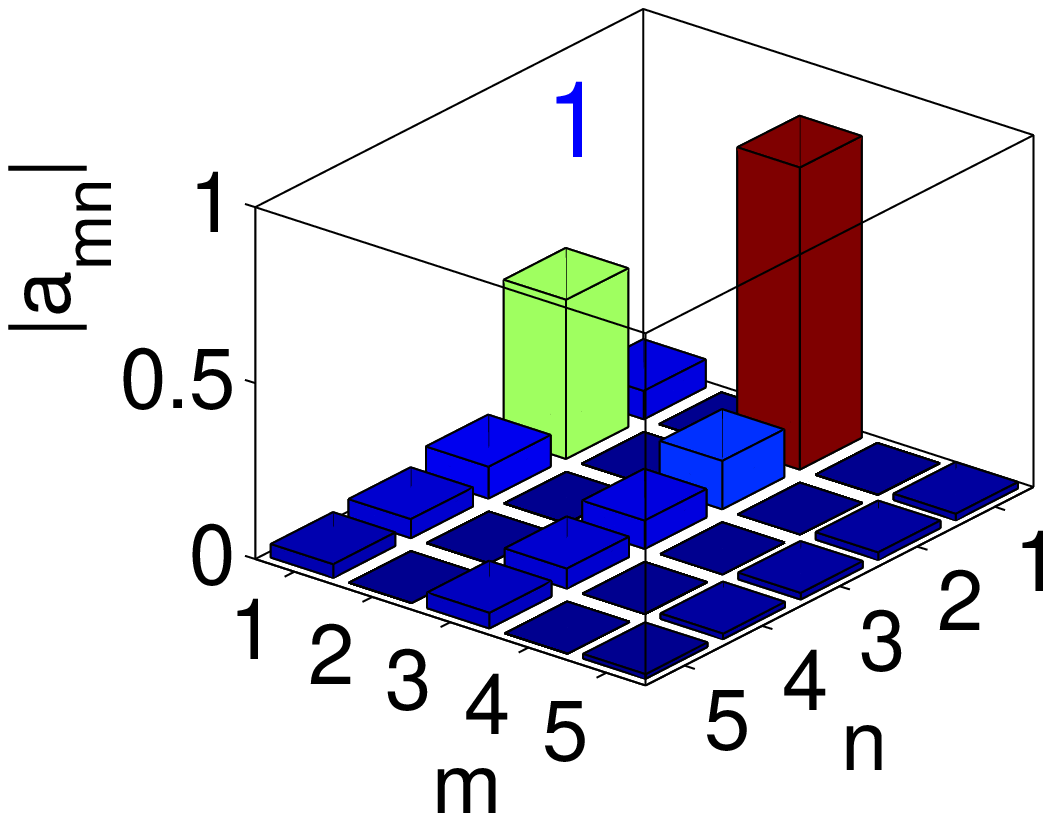}
\includegraphics[height=5cm,width=5cm,clip=]{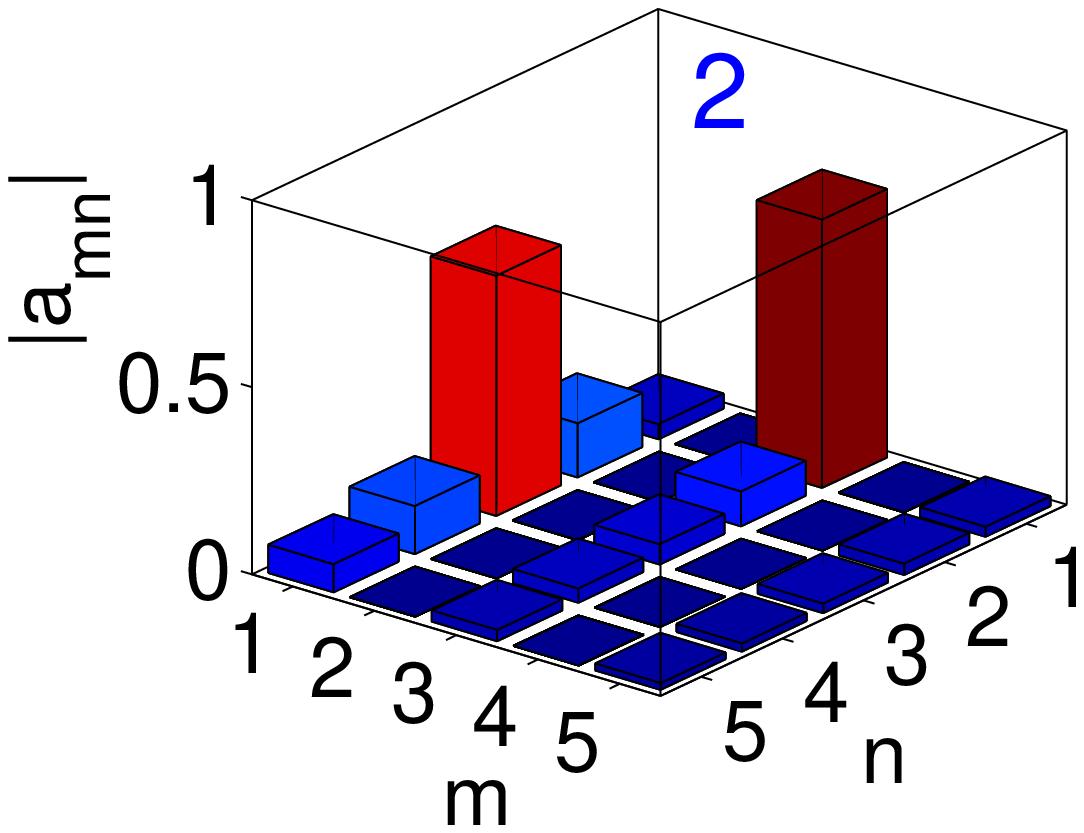}
\caption{(Color online) Expansion of BSC functions in $\Pi$-shaped
waveguide shown in Fig. \ref{fig14} over the
igenfunctions (\ref{psimn}).}
\label{fig15}
\end{figure}
Fig. \ref{fig13}  shows the coefficients of the modal expansion of the BSCs in the $Z$ waveguide
with $L=2$ over the eigenfunctions of the bridge $\psi_{mn}(x,y)$. One can see
that only the first BSC can be approximately described by the Friedrich-Wintgen two-level approach.
Similar expansions for the BSC in the $\Pi$-shaped
waveguide shown in Fig. \ref{fig15} demonstrate that the two-level approximation is applicable.
Moreover one can see that only the eigenfunctions $\phi_m(x)$ symmetrical in the $x$-axis
participate in the BSC which complies the symmetry of the waveguide.
\section{Summary}

We considered electron transmission in two types of double-bend waveguides, namely
$Z$- and $\Pi$-shaped waveguides. The waveguides differ in
the sequence of the chirality of the bends. In the $\Pi$-shaped waveguide the bends have the same chirality
 with respect to the direction
of the flow, while in $Z$-shaped waveguide the bends have opposite chirality. It is demonstrated that because
of the difference in chirality
the vortices near the bends have different current circulation. That explains the difference
in conductance in those two types of waveguide. Alternatively, the origin of resonant peaks in the double-bend
waveguides is explained through the approach of the effective non-Hermitian Hamiltonian.
It is shown that although the central element (bridge)
in both types of waveguide is identical the difference in conductance arises from the structure of the coupling matrix (\ref{Wmn}) which
accounts for the coupling between the bridge and attached waveguides. Moreover the effective Hamiltonian approach
with the coupling matrix (\ref{Wmn}) explains narrowing of resonant peaks with enlarging of the bridge length $L$.
The central result of the paper, however, is the electron localization between the bends thanks to bound states in the continuum (BSC)
with discrete energies embedded in the propagation band of the waveguide.
The BSCs like those in Figs. \ref{fig8}, \ref{fig10}, \ref{fig12}, and \ref{fig14}
are shown to be generic to both $Z$- and $\Pi$-shaped waveguides
subject to the static potential of a finger gate. Such double-bend waveguide
could be seen as a transistor with the finger gate voltage controlling the conductance as well as the
resonant widths. It is shown that the finger gate potential can tune the resonant widths
to zero as a result of variation of the coupling constants between the
eigenstates of the bridge and the first channel continuum.
Although it is hardly possible to visualize electron BSCs in microelectronics, they can be
registered as a singularities in the conductance
where the collapse of the Fano resonances occurs as shown in Figs. \ref{fig7} and
\ref{fig11}.
Such features were experimentally observed by
Lepetit {\it et al} \cite{Lepetit,Lepetit1} for a dielectric resonator positioned in a
microwave waveguide. We speculate that employing the equivalence between quantum waveguides and
microwave systems with TM electromagnetic modes \cite{Stockmann}
it could be possible to directly observe the BSCs in microwave double-bend waveguides.\\

{\bf Acknowledgments}
This work is supported by the Russian Science Foundation
through Grant No.14-12-00266.


\end{document}